\documentclass[mathpazo]{cicp}

\usepackage{amsmath}
\usepackage{enumerate}
\usepackage{multirow}
\usepackage{graphicx,subfigure}

\begin{document}
\title{A positivity-preserving scheme for the simulation of streamer discharges in non-attaching and attaching gases}


 \author[Zhuang C. et.~al.]{Chijie Zhuang,
       Rong Zeng}
 \address{State Key Lab of Power Systems, Department of Electrical Engineering, Tsinghua University, Beijing 100084, China.}
 \emails{{\tt chijie@tsinghua.edu.cn} (C.~Zhang), {\tt zengrong@tsinghua.edu.cn} (R.~Zeng)}

\begin{abstract}
Assumed having axial symmetry, the streamer discharge is often described by a fluid model in cylindrical coordinate system, which consists of convection dominated (diffusion) equations with source terms, coupled with a Poisson's equation. Without additional care for a stricter CFL condition or special treatment to the negative source term, popular methods used in streamer discharge simulations, e.g., FEM-FCT, FVM, cannot ensure the positivity of the particle densities for the cases in attaching gases. By introducing the positivity-preserving limiter proposed by Zhang and Shu \cite{ppl} and Strang operator splitting, this paper proposed a finite difference scheme with a provable positivity-preserving property in cylindrical coordinate system, for the numerical simulation of streamer discharges in non-attaching and attaching gases. Numerical examples in non-attaching gas (N$_2$) and attaching gas (SF$_6$) are given to illustrate the effectiveness of the scheme.
\end{abstract}

\ams{65Z05,65M06, 68U20}
\keywords{WENO finite difference, positivity-preserving, streamer discharge, numerical simulation}

\maketitle

\section{Introduction}
\label{sec1}
As the initial stage of various
electrical discharges such as sparks and lightnings, streamer discharges
happen in natural environment and many industrial applications everyday.
Great efforts have been taken for the experimental study of
streamer discharges over several decades \cite{ebert_review}. However, due to the lack of rigorous measurement methods, the existing experiment
data are still insufficient to build a clear picture of streamer discharges, which made numerical simulations an important auxiliary tool to predict detailed physical quantities in the discharge channel.
A better understanding on the physics of streamer formation and propagation
may be achieved by comparing these numerical predictions with experimental observations.

The most frequently used model to describe streamer discharges is the fluid model,
which consists of the particle density continuity
equations (which are convection-dominated equations with source terms) coupled
with a Poisson's equation with axial symmetries:
\begin{eqnarray}
 & \frac{\partial n_e}{\partial t} + \frac{1}{r}\frac{\partial (r {v_{er}}n_e)}{\partial r}+\frac{\partial (v_{ez}n_e)}{\partial z}-\frac{ D_r}{r}\frac{\partial }{\partial r}(r\frac{\partial n_e}{\partial r}) - D_z\frac{\partial^2 n_e }{\partial z^2}=(\alpha-\eta)n_e |\vec v_e|, \label{eq1.1}\\
 & \frac{\partial n_p}{\partial t} + \frac{1}{r}\frac{\partial (r{v_{pr}}n_p)}{\partial r}+\frac{\partial (v_{pz}n_p)}{\partial z}= \alpha n_e |\vec v_e|, \label{eq1.2}\\
 & \frac{\partial n_n}{\partial t} + \frac{1}{r}\frac{\partial (r{v_{nr}}n_p)}{\partial r}+\frac{\partial (v_{nz}n_p)}{\partial z}= \eta n_e |\vec v_e|, \label{eq1.3}\\
 & \frac{1}{r}\frac{\partial }{\partial r}(r \varepsilon_0 \frac{\partial U}{\partial r}) + \frac{\partial }{\partial z}(\varepsilon_0 \frac{\partial U}{\partial z})= e_0 ( n_e + n_n - n_p),\\
 & \vec E = (E_r, E_z)^T = -(\frac{\partial U}{\partial r}, \frac{\partial U}{\partial z})^T,~|\vec E|=\sqrt{E_r^2+E_z^2},\\
 & \vec v_{e,p,n} = (v_{(e,p,n)r},v_{(e,p,n)z})^T= \mu_{e,p,n}(|\vec E|) \vec E,~|\vec v_e|=\sqrt{v_{er}^2+v_{ez}^2},
\end{eqnarray}
where $t$ denotes time, $r \in [0, a_1]$, $z\in[b_1, b_2]$, $a_1>0$, and $b_1, b_2\in \mathbb{R}$; $n_{e,p,n}$ are the densities of charged particles, $\mu_{e,p,n}$ are the movability coefficient; $\vec v_{e,p,n}$ is the drift velocity; $D_r$ and $D_z$ are the diffusion coefficients, the index $e$, $p$, $n$ stand for electrons, positive ions, negative ions, respectively. $U$ and $\vec E$ are the electrical potential and electric field, respectively; $\varepsilon_0$ is the dielectric coefficient in air; $e_0$ is the unit charge of an electron. $\alpha$ and $\eta$ are measured by experiments and $\alpha > 0$, $\eta > 0$. They are functions of $|\vec E|/N$, i.e., electric field strength $|\vec E|$ divided by the neutral gas number density $N$, see Fig. 1 for an example; in addition,
there exists such critical values $\mbox{E}_1$ for each gas
that
\begin{equation} \label{eq7}
    \begin{cases}
         \alpha \le \eta, &  \text{if $\lvert \vec E\rvert \le \text{E}_1$};\\
         \alpha > \eta, &    \text{if $\lvert \vec E\rvert > \text{E}_1 $}.
    \end{cases}
\end{equation}

By Eq~(\ref{eq7}), strictly speaking, the source term in Eq (\ref{eq1.1}) may be either negative or positive for both non-attaching and attaching gases.
However, when the applied voltage is near or a little more than the breakdown voltage, for non-attaching gas,
$\alpha-\eta$ is positive everywhere in the discharge domain; however, $\alpha-\eta\ll 0$ still exists for attaching gases, which leads to a negative source term in Eq (\ref{eq1.1}).

\begin{figure}[!h]
\centering
\subfigure[air, a non-attaching gas]{ \label{fig:subfig:a} 
\includegraphics[width= 0.485 \textwidth]{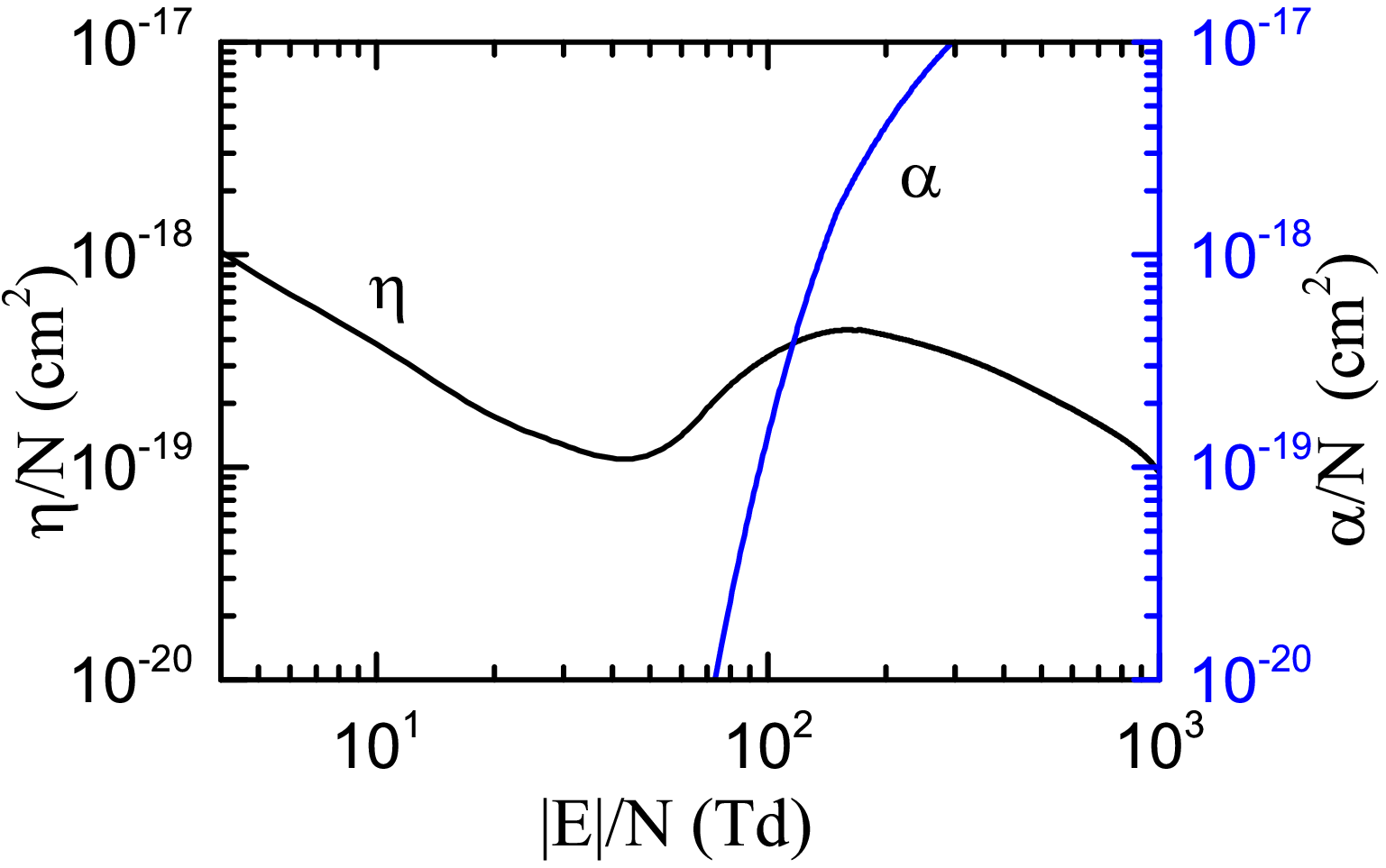}}
\subfigure[SF$_6$, an attaching gas]{
\label{fig:subfig:b} 
\includegraphics[width= 0.485 \textwidth]{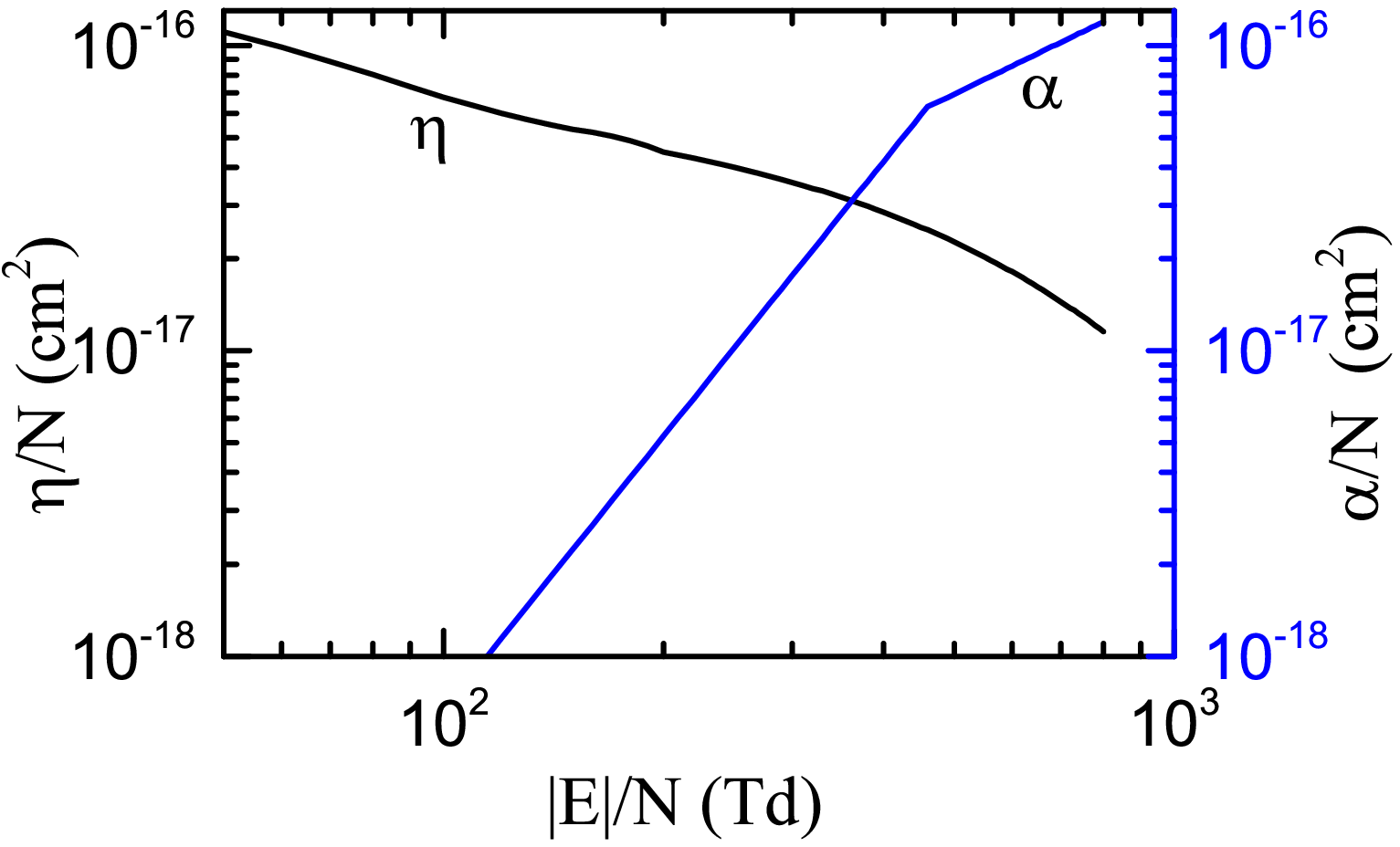}}
\caption{$\alpha$ and $\eta$ for different gases \mbox{(}$1 \mbox{Td}=10^{-17}\mbox{V}\cdot\mbox{cm}^2$, $N\approx 2.446\times 10^{19} \mbox{cm}^{-3}$\mbox{)}}
\end{figure}

For several decades, researchers to the paradigm of streamer discharge simulations have been focusing on the solution of the convection dominated particle density continuity equations, especially on the discretization of the convection term.
Due to the ionization and charge accumulation effect, the particle density profile at the streamer's head is very sharp. Thus, high order linear schemes solving convection dominated Eq (1)-(3) may fail due to numerical oscillations, while first order schemes may add too much
numerical diffusion and smooth the particle density gradient.

The above problems were overcome by nonlinear schemes. The flux-corrected transport (FCT) algorithm \cite{fct1,fct2,fct3}, which used high order ($\ge 2$) solution as much as possible and uses anti-diffusion term to limit the solution in the physical range, were introduced to the field of streamer simulations,  e.g., FDM-FCT by Morrow \cite{morrow1}, Dhali and Williams \cite{dhali1985, dhali1987}, and FEM-FCT used by Morrow and Georghiou \cite{geo}, Min \cite{min}. The finite volume (FV) schemes, e.g., the FV scheme based on Koren's limiter by Ebert \cite{ebert}, the MUSCL scheme used by Papageorghiou et al.\cite{Papageorghiou}, the ULTIMATE QUICKEST used by Bessieres et al.\cite{Bessieres} and Pancheshnyi et al.\cite{Pancheshnyi}, gradually become popular since 2000. The schemes mentioned above are generally free of numerical oscillations. However, a question raised for the above mentioned algorithms: Can they always ensure the particle densities to be positive especially for cases in attaching gases, or at what price do they preserve the positivity?

FCT and the about mentioned FVM are in principle monotone thus positivity preserving, thanks to the limiters or similar strategies they used. However, it is also due to the limiters that these schemes reduce to first order accuracy at local extremes. In addition, FVM schemes like MUSCL may be diffusive for long term simulations of streamer discharges, which makes the streamer charge propagates more rapidly \cite{bourncomp}. Further more,  the extra stricter time step restriction to ensure the positivity in the existence of negative source terms has not been carefully considered in previous literatures. Some previous researchers using FVM for streamer discharge simulations used the limiters in cylindrical coordinate system, and used the physical variable instead of the conservative variable when constructing the numerical flux, which may make the schemes not positivity-preserving near the origin $r=0$ under the mild CFL condition $\alpha \lambda \leq 1$ for a single Euler forward step, see an example in the appendix.

In streamer discharge simulations, a non-oscillatory, positivity preserving solution to the particle density is a basic requirement. Though negative numerical solutions do not blow up the simulations, however, the convection-diffusion equations which describe the charges' motion are coupled with the Poisson's equation, i.e., the charge densities are the input of the Poisson's equation, which determines the electric field distribution in space. When the solution of charge densities are negative, considering their physical effect to the electric field distribution, the polarity of the charges are changed, e.g, positive charges with negative densities, is equivalent to negative charges. In addition, when the charges with negative densities accumulate, their distortion to the electric field may become larger, which may even drive the charges to drift along a wrong direction. Some researchers added same amount of electrons and positive ions to keep the density of electrons always above zero. However, this would alter the reaction terms. Thus a positivity-preserving scheme for streamer discharge simulations is highly desired, especially for the streamer discharge simulations in attaching gases. By introducing the positivity-preserving limiter proposed by Zhang and Shu \cite{ppl} and Strange operator splitting \cite{strang}, this paper proposed a high order WENO finite difference scheme with a provable positivity-preserving property in cylindrical coordinate system, for the numerical simulation of streamer discharges in non-attaching and attaching gases.

This paper is organized as follows. We first consider a 1-dimensional positivity-preserving WENO finite difference scheme for the convection equations in cylindrical coordinate system without source terms and illustrate the main idea to preserve the positivity. A sufficient condition for convection problems to preserve the positivity and the related limiter to enforce this condition is given. After that, we consider the cases with diffusion and source terms, and give the additional CFL condition required to guarantee the positivity. Then the scheme is extended to 2-dimension. Numerical simulations of streamer discharges in non-attaching gas (N$_2$) and attaching gas (SF$_6$) are given to show the effectiveness of the scheme. Finally, we draw some conclusions.

\section{1-dimensional positivity-preserving WENO finite difference scheme for convection equations with axial symmetry}
Taking the governing equation of electrons for example, multiplying Eq (1.1) by $r$, we get
\begin{equation} \label{eq8}
  \frac{\partial (rn_e)}{\partial t}+ \frac{\partial (r {v_{er}}n_e)}{\partial r}+\frac{\partial (rv_{ez}n_e)}{\partial z}- \frac{\partial }{\partial r}(rD_r\frac{\partial n_e}{\partial r}) - \frac{\partial }{\partial z}(rD_z\frac{\partial n_e}{\partial z})=r(\alpha-\eta)n_e |\vec v_e|.
  \end{equation}

To illustrate the main idea of the positivity-preserving WENO finite difference scheme for Eq (\ref{eq8}), we start from 1-dimensional cases.
\subsection{WENO finite difference scheme for convection equations: monotone cases}
We first consider the following 1D case:
\begin{equation} \label{eq1d}
  \frac{\partial (ru)}{\partial t}+ \frac{\partial \left(rf(u) \right)}{\partial r}=0, ~ r\in [0, a_1],~a_1 > 0,~t\geq 0. \end{equation}
In addition, we assume $f'(u)\geq 0$ and $f(0)=0$. In our case of Eq (\ref{eq8}), $f(0)=0$ is satisfied.

For simplicity, we choose the spatial cell size $\triangle r=\frac{a_1}{K}$ for $K$ being a positive integer, and denote cell $i$ as $[r_{i-\frac{1}{2}}, r_{i+\frac{1}{2}}]$, where $r_i$ is the center of the cell $i$, $r_{i+\frac{1}{2}}=r_i+ \frac{1}{2}\triangle r$ and $r_{i-\frac{1}{2}}=r_i- \frac{1}{2}\triangle r$.
A finite difference scheme for Eq (\ref{eq1d}) is given
\begin{equation}\label{eq10}
\frac{\mbox{d}(r_i u_i)}{\mbox{d}t}+ \frac{1}{\triangle r}\left(\widehat {(rf)}_{r_{i+\frac{1}{2}}} - \widehat {(rf)}_{r_{i-\frac{1}{2}}}\right) = 0.
\end{equation}
Eq (\ref{eq10}) is $k$-th order accurate in space if
\begin{equation}\label{eq11}
\frac{1}{\triangle r}\left(\widehat {(rf)}_{r_{i+\frac{1}{2}}} - \widehat {(rf)}_{r_{i-\frac{1}{2}}}\right) = \frac{\partial(rf)}{\partial r}\vert_{r=r_i}+ O(\triangle r^k).
\end{equation}
If there exists a polynomial $h(r)$ such that
\begin{equation}\label{eq12}
  rf(r) = \frac{1}{\triangle r}\int_{r-\frac{1}{2}\triangle r}^{r+\frac{1}{2}\triangle r} h(\xi)\mbox{d}\xi,
\end{equation}
or in other word, the point value $r_if(u_i)$ is the  average of a polynomial $h(r)$ over the domain $[r_{i-\frac{1}{2}}, r_{i+\frac{1}{2}}]$, then
\begin{equation}
  \frac{\partial (rf(u))}{\partial r}\arrowvert_{r=r_i}= \frac{1}{\triangle r}\left(h(r_{i+\frac{1}{2}})-h(r_{i-\frac{1}{2}})\right).
\end{equation}
So what we need to do to achieve Eq (\ref{eq11}) is to use
\begin{eqnarray} \label{eq14}
  \widehat{(rf)}_{r_{i+\frac{1}{2}}} = h(r_{i+\frac{1}{2}}) + O(\triangle r^k),~~ \widehat{(rf)}_{r_{i-\frac{1}{2}}} = h(r_{i-\frac{1}{2}}) + O(\triangle r^k).
\end{eqnarray}

From Eq (\ref{eq12}), the point value $r_i f(u_i)$ is the average of $h(r)$ over the domain $[r_{i-\frac{1}{2}}, r_{i+\frac{1}{2}}]$, i.e., $\overline {{h}_i} = r_if(u_i)$. In order to achieve Eq (\ref{eq14}), we use the average value of $h(r)$, i.e., $\overline {h}_i$ and the average of cell $i$'s neighbors, to reconstructed the point value $h_{i+\frac{1}{2}}$ at the cell interface $i+\frac{1}{2}$. WENO reconstruction is a good choice for such a reconstruction \cite{weno-shu-review}.

The explicit form of high order WENO schemes can be found in, e.g., \cite{weno-shu-review}. For the third order WENO scheme, $h_{i+\frac{1}{2}}^-$ can be reconstructed as follows :
\begin{subequations}
  \begin{eqnarray}
  &  h_{i+\frac{1}{2}}^{-(0)}=\frac{1}{2} \overline {h}_{i}+ \frac{1}{2} \overline {h}_{i+1},~~h_{i+\frac{1}{2}}^{-(1)}=-\frac{1}{2} \overline {h}_{i-1} + \frac{3}{2} \overline {h}_i; \\
  &  \beta_0 = (\overline {h}_{i+1}-\overline {h}_i)^2, ~~~~~~\beta_1 = (\overline {h}_{i}-\overline {h}_{i-1})^2;\\
  &  \alpha_0 = \frac{2}{3}\frac{1}{(\varepsilon+\beta_0)^2},~\alpha_1 = \frac{1}{3}\frac{1}{(\varepsilon+\beta_1)^2},~\varepsilon \mbox{ is a small constant, e.g,}~\varepsilon=10^{-6};\\
  &  \omega_0 = \frac{\alpha_0}{\alpha_0+\alpha_1},~~~~~~ \omega_1 = \frac{\alpha_1}{\alpha_0+\alpha_1};\\
  &  h_{i+\frac{1}{2}}^-=\omega_0 h_{i+\frac{1}{2}}^{-(0)} + \omega_1 h_{i+\frac{1}{2}}^{-(1)}.
  \end{eqnarray}
\end{subequations}

Summarize the WENO scheme for Eq (\ref{eq1d}):
\begin{enumerate}
  \item at the time level $n$, obtain the cell average of $h$ on cell $i$ by $\overline h_i^n = r_i f(u_i)$;
  \item use WENO reconstruction based on $\overline h_i^n$ and $\overline h_j^n$, where cells $j$ are in the neighborhood of cell $i$, to construct the point value at $r_{i+\frac{1}{2}}$, and denote it by $h_{i+\frac{1}{2}}^-$;
  \item set the flux $\widehat{(rf)}_{i+\frac{1}{2}}=h_{i+\frac{1}{2}}^-$;
  \item solve Eq (\ref{eq10}) to obtain the point value $u_i^{n+1}$ at time level $n$+1.
\end{enumerate}

Set $\lambda = \frac{\triangle t}{\triangle r}$. Discretizing Eq (2.3) in time by forward Euler method, the scheme reads
\begin{eqnarray} \label{eq210}
   r_i u_i^{n+1}= r_i u_i^{n} - \lambda (\widehat{rf}_{i+\frac{1}{2}} - \widehat{rf}_{i-\frac{1}{2}} ) = r_i u_i^{n}  - \lambda ({h}^-_{i+\frac{1}{2}} - {h}^-_{i-\frac{1}{2}}).
\end{eqnarray}

$\bf {Remarks}$: In scheme (\ref{eq210}), the point value $r_if(u_i)$ rather than $f(u_i)$ is regarded as the cell average. If $f(u_i)$ is regarded as the cell average of an implicit polynomial $h$, i.e., $f(u) = \frac{1}{\triangle r}\int_{r-\frac{1}{2}\triangle r}^{r+\frac{1}{2}\triangle r} h(\xi)\mbox{d}\xi$.
Assume $h_{i+\frac{1}{2}}$ and $h_{i-\frac{1}{2}}$ are reconstructed by $k$-th order ($k\geq 3$) WENO procedure, i.e.,
\begin{equation} \label{eqr1}
  \frac{\mbox{d} f(u)}{\mbox{d} r}|_{r=r_i}=\frac{1}{\triangle r}(h_{i+\frac{1}{2}}-h_{i-\frac{1}{2}}) + O(\triangle r^k),~~k\geq 3.
\end{equation}

On the other hand, by trapezoidal quadrature,
\begin{equation}\label{eqr2}
  f(u_i)=\frac{1}{2}(h_{i+\frac{1}{2}}+h_{i-\frac{1}{2}})+O(\triangle r^2).
\end{equation}

Combining Eq (\ref{eqr1}) and Eq (\ref{eqr2}), a direct calculation shows,
  \begin{eqnarray}
 && \frac{\partial (rf(u))}{\partial r}|_{r=r_i}-\frac{r_{i+\frac{1}{2}}h_{i+\frac{1}{2}}-r_{i-\frac{1}{2}}h_{i-\frac{1}{2}}}{\triangle r} \nonumber\\
&& =\left( f(u_i)-\frac{h_{i+\frac{1}{2}}+h_{i-\frac{1}{2}}}{2}\right)+r_i\left(\frac{\partial f(u)}{\partial r}|_{r=r_i}-\frac{h_{i+\frac{1}{2}}-h_{i-\frac{1}{2}}}{\triangle r}\right) \nonumber \\
&& =O(\triangle r^2),
\end{eqnarray}
the resulted scheme is 2nd order at most even if higher order ($\geq 3$) WENO is used.

\subsection{A sufficient condition for the positivity-preserving property: monotone cases}
Provided $u>0$, for a monotone $f(u)$ which satisfies $0\leq  f'(u)\leq \alpha$ and $f(0)=0$, it's directly forward to get by Taylor's expansion that  $0\leq f(u)\leq \alpha u$, and $  r\left(u - \frac{f(u)}{\alpha}\right) \geq 0$ for any $r>0$.
Eq (\ref{eq210}) reads
 \begin{eqnarray} \label{eq2.10}
  r_i u_i^{n+1} &=& r_i u_i^n + \lambda (h_{i-\frac{1}{2}}^- - h_{i+\frac{1}{2}}^-) \nonumber \\
  &=& r_i\left( u_i^n - \frac{f(u_i)}{\alpha}\right ) + \left(\frac{r_i f(u_i)}{\alpha} - \lambda h_{i+\frac{1}{2}}^-\right) + \lambda h_{i-\frac{1}{2}}^- \nonumber \\
  &=& H_1 + H_2 + H_3,
\end{eqnarray}
where $H_1 = r_i \left(u_i^n - \frac{f(u_i)}{\alpha}\right)\geq 0$, $H_2=\frac{r_i f(u_i)}{\alpha} -\lambda h_{i+\frac{1}{2}}^-$, and $H_3=\lambda h_{i-\frac{1}{2}}^-$.

To preserve the positivity of $u$, it suffices to ensure $H_2 \geq 0$ and $H_3\geq 0$. Since $r_if(u_i)$ is the cell average of $h(r)$ over cell $i$, by Gauss-Lobbato quadrature, we have
\begin{equation} \label{eq2.11}
  r_if(u_i) = \overline h=\sum_{j=1}^N \omega_jh(r_j) = \omega_N h_{i+\frac{1}{2}}^- + \sum_{j=1}^{N-1}\omega_j h(r_j) ,
\end{equation}
where $\omega_j$ is the quadrature coefficients and $\sum_{j=1}^N \omega_j =1$, $0 < \omega_j < 1$, $\omega_1=\omega_N$.

Assume $h\in [m_1,M_1]$ in cell $i$, and define $h^{-*}=\frac{1}{1-\omega_N}\left(\sum_{j=1}^{N-1} \omega_j h(r_j)\right)=\frac{1}{1-\omega_N}(\overline h - \omega_N h_{i+\frac{1}{2}}^-)$, then $m_1\leq h^{-*}\leq M_1$. By the mean value theorem, there exists an $r^*$ in cell $i$ such that $h(r^*)=h^{-*}$.
Eq (\ref{eq2.11}) reads \begin{equation}
  r_if(u_i) = \overline h = (1-\omega_N) h^{-*}+\omega_N h_{i+\frac{1}{2}}^-.   \label{neweq2.12}
  \end{equation} Plugging Eq (\ref{neweq2.12}) into Eq (\ref{eq2.10}), Eq (\ref{eq2.10}) reads
\begin{equation} \label{eq2.12}
  r_i u_i^{n+1} = H_1 + \frac{(1-\omega_N) h^{-*} + (\omega_N - \alpha \lambda)h_{i+\frac{1}{2}}^-}{\alpha} + H_3 .
\end{equation}

A sufficient condition for scheme (\ref{eq2.12}) to be positivity-preserving is given \cite{thesis, pplsurvey}:
\begin{theorem}
\label{maxmin}
Given positive $u_i^n$, consider a finite difference scheme (\ref{eq210}) or equally scheme (\ref{eq2.12}), associated with the approximation polynomial h,
then $u_i^{n+1} > 0$ if \begin{equation} \label{suf1}
h^{-*} \geq 0,~~ {h}^-_{i+\frac{1}{2}} \geq 0, ~~{h}^-_{i-\frac{1}{2}} \geq 0 \mbox{~~~and~~~} \alpha\lambda \le \omega_N
\end{equation}
provided that the inequalities don't achieve the equality signs at the same time.
\end{theorem}

More theoretical backgrounds about Theorem \ref{maxmin} can be found, e.g, in \cite{thesis, pplsurvey}.

The above sufficient condition for WENO finite difference scheme to ensure the positivity, largely depends on the fact that the point value $r_i f(u_i)$ itself is the cell average of an implicitly existing polynomial $h(r)$, which makes it possible for the point value to be expressed by Gauss-Lobatto quadrature whose quadrature points include the two cell ends, which finally constructs a connection between the point value $r_if(u_i)$ and the fluxes ${h}^-_{i\pm \frac{1}{2}}$ at cell interfaces.

{\bf Remarks:} Without the assumption $f(0)=0$, $H_1$, $H_2$ and $H_3$ would read
\begin{subequations}
  \begin{eqnarray}
    H_1 = r_i \left(u_i^n - \frac{f(u_i)-f(0)}{\alpha}\right)\geq 0,\\
    H_2=\frac{r_i f(u_i)-r_i f(0)}{\alpha} -\lambda (h_{i+\frac{1}{2}}^--{r_i f(0)}),\\
    H_3=\lambda (h_{i-\frac{1}{2}}^--{r_i f(0)}).
  \end{eqnarray}
\end{subequations}
 Since $h(r)$ approximates $rf$ as a whole rather than $f$, thus $rf(u)\geq r_if(0)$ is not guaranteed over the domain $[r_{i-\frac{1}{2}}, r_{i+\frac{1}{2}}]$ although $f(u)\geq f(0)$, therefore we are not able to get a sufficient condition like Eq (\ref{suf1}) using the strategy (\ref{eq2.11}). This may be cured by a different definition of cell average, i.e.,
\begin{equation}
f = \frac{1}{\int_{r-\frac{1}{2}\triangle r}^{r+\frac{1}{2}\triangle r}\xi\mbox{d}\xi}\int_{r-\frac{1}{2}\triangle r}^{r+\frac{1}{2}\triangle r}\xi h(\xi)\mbox{d}\xi.
 \end{equation}
By this way, $h(r)$ approximates $f(r)$ rather than $rf(r)$. However, this leads to a new finite volume scheme and we leave it as a future work.

{\bf Remarks:} For the cases $-\alpha \leq f'(u)<0$ and $f(0)=0$, we have $\alpha u + f(u)\geq0$. Eq (\ref{eq2.10}) reads
\begin{eqnarray} \label{remark}
 r_i u_i^{n+1} &=& r_i u_i^n + \lambda (h_{i-\frac{1}{2}}^+ - h_{i+\frac{1}{2}}^+) \nonumber \\
  &=& r_i\left( u_i^n + \frac{f(u_i)}{\alpha}\right ) + \left(\lambda h_{i-\frac{1}{2}}^+ -\frac{r_i f(u_i)}{\alpha}\right) - \lambda h_{i+\frac{1}{2}}^+.
\end{eqnarray}

Denote $h^{+*}=\frac{1}{1-\omega_1}(\overline h- \omega_1 h_{i-\frac{1}{2}}^+)$. Similarly, a sufficient condition for Eq (\ref{remark}) to preserve the positivity of $u$ is given:
\begin{theorem}
Given positive $u_i^n$, consider a finite difference scheme (\ref{remark}), associated with the approximation polynomial h,
then $u_i^{n+1} > 0$ if \begin{equation}  \label{eq219}
h^{+*} \leq 0,~~ {h}^+_{i+\frac{1}{2}} \leq 0, ~~{h}^+_{i-\frac{1}{2}} \leq 0 \mbox{~~~and~~~} \alpha\lambda \le \omega_1
\end{equation}
provided that the inequalities don't achieve the equality signs at the same time.
\end{theorem}

\subsection{A linear scaling limiter}\label{ppl}
The sufficient condition (\ref{suf1}) can be enforced by a linear scaling limiter \cite{scalarppl}. Assuming $\overline h \in [m, +\infty)$ with $m\geq 0$, it suffices to apply the following limiter to make $\widehat{h}_{i+\frac{1}{2}}^-\geq m$ and $\widehat{h}^{-*}\geq m$:
\begin{equation} \label{eq218}
\widehat{h}(r) = \theta ( h(r) - \overline h) + \overline h,~~\theta = \min\left\{\frac{\overline h-m}{\overline h-q_{\min}},1\right\}.
\end{equation}
where $q_{\min}=\min(h_{i+\frac{1}{2}}^-, h^{-*})$. Then use $\widehat {h}_{i+\frac{1}{2}}^-$ instead of $h_{i+\frac{1}{2}}^-$ in Eq (\ref{eq210}). To enforce sufficient condition (\ref{suf1}), we set $m=0$.

${\bf Remarks:}$ Eq (\ref{eq218}) is for the cases where $f'(u)\geq 0$. For the cases where $f'(u)<0$ and $\overline h \in (-\infty, M]$ with $M\leq 0$, the limiter reads  $\widehat{h}(r) = \theta ( h(r) - \overline h) + \overline h,~~\theta = \min\left\{\frac{\overline h-M}{\overline h-q_{\max}},1\right\}$, where $q_{\max}=\max(h_{i-\frac{1}{2}}^+, h^{+*})$. Then $\max(\widehat{h}_{i-\frac{1}{2}}^+, \widehat{h}^{+*}) \leq M$ after limiting. To enforce the sufficient condition (\ref{eq219}), we set $M=0$.

${\bf Remarks:}$ As already stated by X. Zhang \cite{ppl}, $h(r)$ would have a smaller minimum than $rf \in [m, +\infty)$. If we enforce $h(r) \in[m, +\infty)$, order degradation would occur. Fortunately, the positivity-preserving limiter only needs to be turned on when the positivity is violated, while in regions with strictly positive solution, the limiter can be turned off and does not cause order degradation even at positive local extremes.

\subsection{WENO finite difference scheme for convection equations: general cases}
On general occasions that $f(u)$ is not locally monotone over the stencil,e.g., $I_{i+\frac{1}{2}}$, which is for a flux construction at the interface $r_{i+\frac{1}{2}}$, the following flux splitting is performed:
\begin{equation}  \label{eq16}
  p_{+}^{i+\frac{1}{2}} = \frac{1}{2}r\left(u+\frac{f(u)}{\alpha_{i+\frac{1}{2}}}\right),~~~p_{-}^{i+\frac{1}{2}}=\frac{1}{2}r\left(u-\frac{f(u)}{\alpha_{i+\frac{1}{2}}}\right),
\end{equation}
where $\alpha_{i+\frac{1}{2}}$ is the local maximum of ${|f'(u)|}$ over the stencil. The superscript ${i+\frac{1}{2}}$ is used to clarify that the splitting is related to the interface $r_{i+\frac{1}{2}}$. We have
\begin{subequations}
\begin{eqnarray}
&  r_iu_i=\frac{1}{2}r_i\left(u_i+\frac{f(u_i)}{\alpha_{i+\frac{1}{2}}}\right)+ \frac{1}{2}r_i\left(u_i-\frac{f(u_i)}{\alpha_{i+\frac{1}{2}}}\right)= p_{i,+}^{i+\frac{1}{2}}+p_{i,-}^{i+\frac{1}{2}}, \\
&  r_if_i= \alpha_{i+\frac{1}{2}}\left(\frac{1}{2}r_i(u_i+\frac{f(u_i)}{\alpha_{i+\frac{1}{2}}})- \frac{1}{2}r_i(u_i-\frac{f(u_i)}{\alpha_{i+\frac{1}{2}}})\right)=\alpha_{i+\frac{1}{2}}(p_{i,+}^{i+\frac{1}{2}}-p_{i,-}^{i+\frac{1}{2}}).
\end{eqnarray}
\end{subequations}
Under the assumption $f(0)=0$, we still have $p_{+}^{i+\frac{1}{2}}(0)=p_{-}^{i+\frac{1}{2}}(0)=0$. In addition, $\frac{\partial p_{\pm}^{i+\frac{1}{2}}}{\partial u} \geq 0$, which implies $p_{\pm}^{i+\frac{1}{2}}(u)\geq p_{\pm}^{i+\frac{1}{2}}(0)= 0$ for all $u\geq 0$.

At time level $t^n$, for each fixed interface $r_{i+\frac{1}{2}}$, the procedure to reconstruct the flux $\widehat{rf}_{i+\frac{1}{2}}$ is given as follows:
\begin{enumerate}
  \item choose $\alpha_{i+\frac{1}{2}}=\max_j {|f'(u)|}$, for all $j$ in the stencil $I_{i+\frac{1}{2}}$.
  \item  obtain the point values $p_{j,+}^{i+\frac{1}{2}}$ and $p_{j,-}^{i+\frac{1}{2}}$ by Eq (\ref{eq16}) for all $j$ in the stencil, and set the cell averages $\overline q_{j,+}^{i+\frac{1}{2}} = p_{j,+}^{i+\frac{1}{2}}$ and $\overline q_{j,-}^{i+\frac{1}{2}}= p_{j,-}^{i+\frac{1}{2}}$, respectively.
  \item reconstruct the point values at the interface $r_{i+\frac{1}{2}}$, i.e., $q_{i+\frac{1}{2},+}^l$ and $q_{i+\frac{1}{2},-}^r$, by WENO reconstructions,  based on the cell averages $\overline q_{j,+}^{i+\frac{1}{2}}$ and $\overline q_{j,-}^{i+\frac{1}{2}}$, respectively; the superscript $l$ and $r$ mean they are the value at the left side or right side of the interface $r_{i+\frac{1}{2}}$ respectively.
  \item obtain $\widehat{rf}_{i+\frac{1}{2}}=\alpha_{i+\frac{1}{2}}(q_{i+\frac{1}{2},+}^l- q_{i+\frac{1}{2},-}^r)$.
\end{enumerate}

The finite difference scheme for Eq (\ref{eq1d}) is
\begin{equation} \label{eq2.20}
  r_i u_i^{n+1} = r_i u_i^n - \lambda \left(\widehat{(rf)}_{i+\frac{1}{2}} - \widehat{(rf)}_{i-\frac{1}{2}}\right).
\end{equation}

\subsection{A sufficient condition for the positivity-preserving property: general cases}
Given a general function $f(u)$, in a WENO finite difference scheme listed above, neither the point value $r_i f(u_i)$ nor $r_i u_i$ is a cell average of a single polynomial. However, they can be regarded as linear combinations of the cell averages of two polynomials, which make us possible to construct a connection between the point values  and the fluxes $\widehat {rf}_{i\pm\frac{1}{2}}$ by Gauss-Lobatto quadrature. Scheme (\ref{eq2.20}) reads
\begin{eqnarray}  \label{eq19}
  r_iu_i^{n+1}&=&r_i u_i^{n} -\lambda \left(\widehat{rf}_{i+\frac{1}{2}}-\widehat{rf}_{i-\frac{1}{2}}\right) = \frac{1}{2} (r_i u_i^{n}+ r_i u_i^{n})-\lambda \left(\widehat{rf}_{i+\frac{1}{2}}-\widehat{rf}_{i-\frac{1}{2}}\right)  \nonumber\\
              &=&\frac{1}{2}\bigg(\overline q_{i,+}^{i+\frac{1}{2}} +\overline q_{i,-}^{i+\frac{1}{2}}+ \overline q_{i,+}^{i-\frac{1}{2}} +\overline q_{i,-}^{i-\frac{1}{2}}\bigg)\nonumber \\ &&-\lambda \bigg(\alpha_{i+\frac{1}{2}}(q_{i+\frac{1}{2},+}^l- q_{i+\frac{1}{2},-}^r) - \alpha_{i-\frac{1}{2}}(q_{i-\frac{1}{2},+}^l- q_{i-\frac{1}{2},-}^r)\bigg) \nonumber \\
              &=&Q_1 + Q_2 ,
\end{eqnarray}
where
\begin{subequations}
  \begin{eqnarray}
  &  Q_1=\frac{1}{2}\overline q_{i,-}^{i+\frac{1}{2}} +\lambda \alpha_{i-\frac{1}{2}}q_{i-\frac{1}{2},+}^l +\frac{1}{2}\overline q_{i,+}^{i+\frac{1}{2}}-\lambda \alpha_{i+\frac{1}{2}}q_{i+\frac{1}{2},+}^l, \\
  & Q_2=\frac{1}{2}\overline q_{i,+}^{i-\frac{1}{2}} +\lambda \alpha_{i+\frac{1}{2}}q_{i+\frac{1}{2},-}^r +\frac{1}{2}\overline q_{i,-}^{i-\frac{1}{2}}-\lambda \alpha_{i-\frac{1}{2}}q_{i-\frac{1}{2},-}^r.
  \end{eqnarray}
\end{subequations}

Assume there exists a polynomial $h_1$ of degree $k$ whose cell average on cell $i$ is $\overline q_{i,+}^{i+\frac{1}{2}}$, such that $h_{1,{i+\frac{1}{2}}}=q_{i+\frac{1}{2},+}^l$, $h_{1,{i-\frac{1}{2}}}=q_{i-\frac{1}{2},+}^r$, and $h_1$ is a ($k$+1)-th order accurate approximation to the function $p_+^{i+\frac{1}{2}}$ on the cell if $u$ is smooth. The existence of such a polynomial can be established by interpolation for WENO schemes \cite{weno-shu-review}.
By Gauss-Lobatto quadrature, $\overline q_{i,+}^{i+\frac{1}{2}} = \sum_{j=1}^N \omega_j h_{1j} = (\sum_{j=1}^{N-1}\omega_j h_{1j})+\omega_N q_{i+\frac{1}{2},+}^l$. Similar to the previous section, define
$h_1^* = \frac{1}{1-\omega_N}(\overline q_{i,+}^{i+\frac{1}{2}} - \omega_N q_{i+\frac{1}{2},+}^l)$, then
\begin{equation}
  Q_1= \frac{1}{2}\overline q_{i,-}^{i+\frac{1}{2}} +\lambda \alpha_{i-\frac{1}{2}}q_{i-\frac{1}{2},+}^l+\frac{1}{2}\bigg((1-\omega_N)h_1^* + (\omega_N -2\lambda \alpha_{i+\frac{1}{2}})q_{i+\frac{1}{2},+}^l\bigg).
\end{equation}
To ensure $Q_1\ge 0$, it suffices to provide $q_{i-\frac{1}{2},+}^l \geq 0$, $h_1^* \geq 0$, $\omega_N -2\lambda \alpha_{i+\frac{1}{2}} \geq 0$ and
$q_{i+\frac{1}{2},+}^l \geq 0$.

Similarly, define $h_2^* = \frac{1}{1-\omega_1}(\overline q_{i,-}^{i-\frac{1}{2}} - \omega_1 q_{i-\frac{1}{2},-}^r)$, a sufficient condition for the non-negativity of $Q_2$ is $q_{i+\frac{1}{2},-}^r \geq 0$, $h_2^* \geq 0$, $\omega_1 -2\lambda \alpha_{i-\frac{1}{2}} \geq 0$ and
$q_{i-\frac{1}{2},-}^r \geq 0$.

Let's summarize in the following theorem:
\begin{theorem}
For finite difference scheme (\ref{eq2.20}), given a positive $u_i^n$, $u_i^{n+1}$ is positive if
  \begin{subequations} \label{eq27}
  \begin{eqnarray}
    q_{i-\frac{1}{2},+}^l \geq 0, ~~~h_1^* \geq 0,~~~ q_{i+\frac{1}{2},+}^l \geq 0,  \\
    q_{i-\frac{1}{2},-}^r \geq 0, ~~~h_2^* \geq 0,~~~ q_{i+\frac{1}{2},-}^r \geq 0,
  \end{eqnarray}
  \end{subequations}
provided that $\lambda \max(\alpha_{i+\frac{1}{2}}, \alpha_{i-\frac{1}{2}})\leq \frac{\omega_1}{2}$ and the inequalities don't achieve the equality signs at the same time.
\end{theorem}

In real implementations, the above sufficient condition can be further simplified for different occasions. Assume $f'(u)\geq0$ over stencil $I_{i+\frac{1}{2}}$, and $f(u)$ is not monotone over stencil $I_{i-\frac{1}{2}}$, then due to the fact $\widehat{rf}_{i-\frac{1}{2}}=\alpha_{i-\frac{1}{2}}(q_{i-\frac{1}{2},+}^l- q_{i-\frac{1}{2},-}^r)$ and $r_i u_i = \overline q_{i,+}^{i-\frac{1}{2}} +\overline q_{i,-}^{i-\frac{1}{2}}$, Eq (\ref{eq2.20}) reads
\begin{eqnarray}
    r_iu_i^{n+1}&=&r_i u_i^{n} -\lambda \left(\widehat{rf}_{i+\frac{1}{2}}-\widehat{rf}_{i-\frac{1}{2}}\right) = \bigg(\frac{1}{2} r_i u_i^{n} - \lambda\widehat{rf}_{i+\frac{1}{2}}\bigg) + \bigg(\frac{1}{2} r_i u_i^{n} +\lambda \widehat{rf}_{i-\frac{1}{2}}\bigg)\nonumber\\
              &=&\frac{1}{2}\bigg( r_i u_i -2\lambda\widehat{rf}_{i+\frac{1}{2}}\bigg) + \frac{1}{2}\bigg((\overline q_{i,+}^{i-\frac{1}{2}} +\overline q_{i,-}^{i-\frac{1}{2}}) +2\lambda \alpha_{i-\frac{1}{2}}(q_{i-\frac{1}{2},+}^l- q_{i-\frac{1}{2},-}^r)\bigg) \nonumber \\
              &=&\frac{1}{2}(W_1 + W_2 + W_3),
\end{eqnarray}
where $W_1=r_i u_i -2\lambda\widehat{rf}_{i+\frac{1}{2}}$, $W_2=\overline q_{i,-}^{i-\frac{1}{2}}-2\lambda \alpha_{i-\frac{1}{2}}q_{i-\frac{1}{2},-}^r$, $W_3=\overline q_{i,+}^{i-\frac{1}{2}}+2\lambda \alpha_{i-\frac{1}{2}}q_{i-\frac{1}{2},+}^l$.
It's obvious that $u_i$ is positive if $W_1\geq 0$, $W_2\geq 0$, $W_3\geq 0$, provided at least one inequality does not achieve the equality sign. On this occasion, a sufficient condition to ensure the positivity of $u_i$ is
\begin{equation}
  \underbrace{h_{i+\frac{1}{2}}^- \geq 0, ~h^{-*} \geq 0}_{W_1\geq 0}, ~\underbrace{q_{i-\frac{1}{2},-}^r \geq 0, ~h_2^* \geq 0}_{W_2\geq 0}, ~\underbrace{q_{i-\frac{1}{2},+}^l \geq 0}_{W_3\geq 0},
\end{equation}
provided at least one inequality does not achieve the equality sign, under the CFL condition $\lambda\max{(\alpha_{i-\frac{1}{2}}, \alpha_{i+\frac{1}{2}})}\leq \frac{\omega_1}{2}$.

Other three occasions can be treated similarly, which include: 1), $f'(u)<0$ over stencil $I_{i+\frac{1}{2}}$, and $f(u)$ is not monotone over stencil $I_{i-\frac{1}{2}}$; 2), $f'(u)\geq 0$ over stencil $I_{i-\frac{1}{2}}$, and $f(u)$ is not monotone over stencil $I_{i+\frac{1}{2}}$; 3), $f'(u)<0$ over stencil $I_{i-\frac{1}{2}}$, and $f(u)$ is not monotone over stencil $I_{i+\frac{1}{2}}$. If $f(u)$ is monotone over a stencil, the flux splitting is not necessary.

\section{Generalizations}
\subsection{Generalization to cases with a source term}
Let $S$ be the source term. If $S\ge0$, it's obvious that $u_i^{n+1}\ge0$, if $u_i^n\ge 0$ and the sufficient condition (\ref{eq27}) is satisfied. If $S<0$, the scheme reads, e.g. \cite{ppl},
\begin{eqnarray}
  r_i u_i^{n+1} &=& r_i u_i^n - \lambda \left((\widehat {rf})_{i+\frac{1}{2}} - (\widehat{rf})_{i-\frac{1}{2}}\right) + r_i S(r_i, u_i)\triangle t \nonumber \\  &=& \frac{1}{2}\left(r_i u_i^n - 2\lambda \left((\widehat{rf})_{i+\frac{1}{2}} - (\widehat{rf})_{i-\frac{1}{2}}\right)\right) + \frac{r_i}{2} \left(u_i^n + 2S(r_i, u_i)\triangle t\right). \nonumber
\end{eqnarray}
For the positivity of $u_i^{n+1}$, it suffices to ensure
\begin{subequations}
\begin{eqnarray}
  r_i u_i^n - 2\lambda \left((\widehat{rf})_{i+\frac{1}{2}} - (\widehat{rf})_{i-\frac{1}{2}}\right) \ge 0 ~~\Longrightarrow~~ \triangle t \le \triangle t^0, \label{eq22a}\\
  u_i^n + 2S(r_i, u_i)\triangle t \ge 0 ~~\Longrightarrow~~ \triangle t \le \triangle t^* \label{eq22b}.
\end{eqnarray}
\end{subequations}
How to ensure Eq (\ref{eq22a}) has already been discussed in the previous section. One can choose a time step satisfying the requirement of Eq (\ref{eq22a}) and Eq (\ref{eq22b}), i.e., let $\triangle t\le \min(\triangle t^0, \triangle t^*)$.

In addition, we propose an alternative strategy by Strang operator splitting \cite{strang}. Sometimes, the source term may be a bit stiff, resulting in a small $\triangle t$. On these occasions, the Strang operator splitting together with suitable implicit or exact time integration of the source term can not only achieve a larger time step but also preserve the positivity.

Let $R$ denote the source term, $C$ denote other terms. The symmetric Strang splitting is given:
\begin{enumerate}
  \item solve $\frac{\text{d}u}{\mbox{d}t}=R$ for $\frac{1}{2}\triangle t$ to get $\widehat{u_i}^{n+\frac{1}{2}}$;
  \item solve $\frac{\partial (r_i u_i)}{\partial t}+C=0$ for $\triangle t$ to get $\widehat{u_i}^{n+1}$;
  \item solve $\frac{\mbox{d} u}{\mbox{d} t}=R$ for another $\frac{1}{2}\triangle t$ to get $u_i^{n+1}$.
\end{enumerate}
Also, we have another similar alternative:
\begin{enumerate}
  \item solve $\frac{\partial (r_i u_i)}{\partial t}+C=0$ for $\frac{1}{2}\triangle t$ to get $\widehat{u_i}^{n+\frac{1}{2}}$;
  \item solve $\frac{\mbox{d}u}{\mbox{d}t}=R$ for $\triangle t$ to get $\widehat{u_i}^{n+1}$;
  \item solve $\frac{\partial (r_i u_i)}{\partial t}+C=0$ for another $\frac{1}{2}\triangle t$ to get ${u_i}^{n+1}$.
\end{enumerate}

In our problems, the source terms for Eq (\ref{eq1.1}), (\ref{eq1.2}) and (\ref{eq1.3}) are $R_e =(\alpha-\eta)|\vec v_e|n_e$, $R_p = \alpha |\vec v_e|n_e$ and $R_n=\eta |\vec v_e| n_e$, respectively. Assuming $|\vec v_e|$ remains constant from $t^n$ to $t^{n+\frac{1}{2}}=t^n+\frac{\triangle t}{2}$, the analytic solutions for the reaction terms are given
\begin{subequations}
  \begin{eqnarray}
    \widehat{n_e}^{n+\frac{1}{2}} &=& n_e^{n} \exp\left[(\alpha-\eta)|\vec v_e|\frac{\triangle t}{2}\right].\label{eq3.2a}\\
    \widehat{n_p}^{n+\frac{1}{2}} &=& \begin{cases}
      \frac{\alpha n_e^n}{\alpha-\eta} \left\{\exp\left[(\alpha-\eta)|\vec v_e|\frac{\triangle t}{2}\right]-1\right\}+n_p^{n} &\mbox{if~}\alpha\neq \eta; \cr
      \alpha n_e^{n}|\vec v_e|\frac{\triangle t}{2} +¡¡n_p^{n}& \mbox{if~}\alpha=\eta.
    \end{cases} \label{eq3.2b}\\
    \widehat{n_n}^{n+\frac{1}{2}} &=&
    \begin{cases}
      \frac{\eta n_e^{n}}{\alpha-\eta} \left\{\exp\left[(\alpha-\eta)|\vec v_e| \frac{\triangle t}{2}\right]-1\right\}+n_n^{n} & \mbox{if~}\alpha\neq \eta; \cr
      \eta n_e^{n}|\vec v_e|\frac{\triangle t}{2} +¡¡n_n^{n} & \mbox{if~}\alpha=\eta.
    \end{cases}\label{eq3.2c}
  \end{eqnarray}
\end{subequations}

Provided the initial values at $t^n$ are positive, it's directly forward to show $\widehat{n_e}^{n+\frac{1}{2}} > 0$, $\widehat{n_p}^{n+\frac{1}{2}}\ge n_p^{n} > 0$, $\widehat{n_n}^{n+\frac{1}{2}}\ge n_n^{n}> 0$.

\subsection{Generalization to cases with a diffusion term}
We discretize the diffusion term by 2nd order central finite difference:
\begin{eqnarray}
  r_i u_i^{n+1} &=&r_i u_i^n - \lambda \left((\widehat {rf})_{i+\frac{1}{2}} - (\widehat{rf})_{i-\frac{1}{2}}\right)+ r_i S(r_i, u_i) \nonumber \\
    &~&+{\lambda D_r}\left(r_{i+\frac{1}{2}}\frac{u_{i+1}-u_i}{\triangle r}-r_{i-\frac{1}{2}}\frac{u_i-u_{i-1}}{\triangle r}\right) \nonumber \\
    &=&\frac{1}{4}\left(r_i u_i^n - 4\lambda \left((\widehat{rf})_{i+\frac{1}{2}} - (\widehat{rf})_{i-\frac{1}{2}}\right)\right) + \frac{r_i}{2} \left(u_i^n + 2S(r_i, u_i)\triangle t\right) \nonumber \\
    &~&+\frac{r_i u_i^n}{4}\left(1 - 8\lambda \frac{D_r}{\triangle r} \right)+\lambda D_r\left(\frac{r_{i +\frac{1}{2}}}{\triangle r}u_{i+1}^n+\frac{r_{i -\frac{1}{2}}}{\triangle r}u_{i-1}^n\right). \nonumber
\end{eqnarray}
Since $\left(\frac{r_{i +\frac{1}{2}}}{\triangle r}u_{i+1}^n+\frac{r_{i -\frac{1}{2}}}{\triangle r}u_{i-1}^n\right)\ge 0$, for the positivity of $u_i^{n+1}$, it suffices to ensure
\begin{subequations}\label{eq13}
\begin{eqnarray}
&  r_i u_i^n - 4\lambda \left((\widehat{rf})_{i+\frac{1}{2}} - (\widehat{rf})_{i-\frac{1}{2}}\right) > 0, \\
&  u_i^n + 2S(r_i, u_i)\triangle t \ge 0, \\
&  8D_r\lambda \leq \triangle r.
\end{eqnarray}
\end{subequations}
Condition (\ref{eq13}) is sufficient, but not necessary. Also, one can assign the term $r_i u_i$ for convection, diffusion and source term in different ways for a specific problem to make the allowed $\triangle t$ as large as possible.

\subsection{Generalization to two dimensions}
For simplicity, we only give the case for convection terms. The two dimensional equation is given by
\begin{equation}
  \frac{\partial (ru)}{\partial t}+\frac{\partial \left(rf(u)\right)}{\partial r}+\frac{\partial \left(rg(u)\right)}{\partial z}=0.
\end{equation}
A finite difference scheme is given by, e.g.,
\begin{eqnarray}
 r_i u_{i,j}^{n+1} &=& r_i u_{i,j}^n - \frac{\triangle t}{\triangle r}\left(\widehat{(rf)}_{i+\frac{1}{2},j}-\widehat{(rf)}_{i-\frac{1}{2},j}\right)-\frac{\triangle t}{\triangle z}\left(\widehat{(rg)}_{i,j+\frac{1}{2}}-\widehat{(rg)}_{i,j-\frac{1}{2}}\right) \nonumber \\
 &=&\frac{1}{2}\left(r_i u_{i,j}^n - \frac{2\triangle t}{\triangle r}\left(\widehat{(rf)}_{i+\frac{1}{2},j}-\widehat{(rf)}_{i-\frac{1}{2},j}\right)\right) \nonumber \\&~+&\frac{1}{2}\left(r_i u_{i,j}^n -\frac{2\triangle t}{\triangle z}\left(\widehat{(rg)}_{i,j+\frac{1}{2}}-\widehat{(rg)}_{i,j-\frac{1}{2}}\right)\right).
\end{eqnarray}

By a proper assignment of the term $r_i u_i$, the two dimensional case is split to two
one-dimensional cases and the positivity-preserving limiter can be applied dimension by dimension.

\section{Time integration}
After the space discretization, we get an ODE,
\begin{equation}
  \frac{\mbox{d} u}{\mbox{d}t} = \mbox{L}(u).
\end{equation}
The Total-Variation-Diminishing Runge-Kutta (TVDRK) proposed by Shu is used for time discretizaton \cite{tvd}.
For 2nd order accuracy in time,
\begin{subequations}
  \begin{eqnarray}
  u^{(0)} &=& u^n, \\
  u^{(1)} &=& u^{(0)}+\mbox{L}(u^{(0)})\triangle t, \\
  u^{n+1} &=& \frac{1}{2}u^{n} + \frac{1}{2} \left( u^{(1)} + \mbox{L}(u^{(1)})\triangle t\right).
\end{eqnarray}
\end{subequations}

In our simulations, if Strang splitting is not applied, TVDRK is used for all the parts including convection/diffusion and source terms;
or the TVDRK is only used to solve the convection/diffusion parts and the reaction parts are solved exactly.
Since TVDRK is a convex combination of Euler forward, and the exact integration of the source terms is positivity-preserving, the full scheme is still positivity-preserving.

\section{Whole algorithm for streamer simulations}
The Poisson's equation is discretized by 2nd order central finite difference scheme.
\begin{subequations}
\begin{equation}
 \frac{u_{i+1,j}-2u_{i,j}+u_{i-1,j}}{\triangle r^2}+\frac{u_{i+1,j}-u_{i-1,j}}{2r_i\triangle r}+\frac{u_{i,j+1}-2u_{i,j}+u_{i,j-1}}{\triangle z^2}=\frac{e_0}{\varepsilon_0}\left(n_{e;(i,j)}+n_{n;(i,j)}-n_{p;(i,j)}\right)\label{eq30a}
 \end{equation}
 \begin{equation}
 E_{i,j}=(E_{r;i,j}, E_{z;i,j})^T= \left(\frac{u_{i-1,j}-u_{i+1,j}}{2\triangle r}, \frac{u_{i,j-1}-u_{i,j+1}}{2\triangle z}\right)^T. \label{eq30b}
\end{equation}
\end{subequations}
Eq (\ref{eq30a}) can be solved by FISHPACK, which is based on cyclic reduction and Fast Fourier transform \cite{fishpack}.

At time level $t^n$, given $n_e^n$, $n_p^n$, $n_n^n$, the whole simulation flowchart is as follows:
 \begin{enumerate}
   \item solve Poisson's equation to get the electric field, by Eq (\ref{eq30a}) and (\ref{eq30b});
   \item calculate all the necessary coefficients in Eq (\ref{eq1.1})-(\ref{eq1.3}), i.e., $\alpha$, $\eta$, $v_e$, $v_p$, $v_n$;
   \item use the positivity-preserving scheme described in Section 2 and Section 3 to solve Eq (1.1)-(1.3), either using Strang splitting or not, and get $n_e^{n+1}$, $n_p^{n+1}$, $n_n^{n+1}$.
   \item move to the next time level $t^{n+1}$ and go to step 1.
 \end{enumerate}

 Below we will call the method using Strang splitting with exactly solved reactions the Method I, and call the other one the Method II.

\section{Numerical examples for the positivity-preserving scheme}
We use pure advection problems to test the effectiveness of the positivity-preserving scheme.
\subsection{a case with smooth solutions}
The following problem is solved by fifth order WENO finite difference scheme (WENO5) together with third order TVDRK.
\begin{subequations}
\begin{eqnarray}
 & \frac{\partial u}{\partial t}+ \frac{1}{r}\frac{\partial (ru)}{\partial t} = 0, ~~r\in[a,b],~~a=0.0001,~b= a+1. \\
 & n(r,t=0) = \frac{1}{r}(1.0001+\sin(2\pi(r-a)),~~au(a,t)=b u(b,t).
\end{eqnarray}
\end{subequations}

\begin{table}[ht]
\caption{comparison of errors with and without the limiter when $t=0.5$, with $\triangle t=(\triangle r)^{\frac{5}{3}}$}
\centering
\begin{tabular}{c c c c c| c c c c c }
\hline \hline
\multirow{2}{*}{$\frac{1}{\triangle r}$} &
\multicolumn{4}{c|}{without the limier} &
\multicolumn{4}{c}{with the limiter} \\
\cline{2-5} \cline{6-9}
& $||r_iu_i-r_iu_i^h||_1 $ & order & $||u_i-u_i^h||_1 $ & order & $||r_iu_i-r_iu_i^h||_1 $ & order &  $||u_i-u_i^h||_1 $ & order\\
\hline
{20} & 7.60e-4&      &2.56e-3 &     &               1.10e-3  & & 4.43e-3 & \\
{40} & 2.28e-5 & 5.06& 7.54e-5 &  5.09&                      1.35e-4& 3.03& 6.50e-4 &2.77\\
{80} & 7.05e-7 & 5.02& 2.40e-6 & 4.97&   1.50e-5 & 3.17&6.36e-5 &3.35\\
{160}& 2.20e-8 & 5.00& 7.64e-8 & 4.97 &  2.20e-8  & 9.41&7.64e-8 &9.70\\
{320}& 6.87e-10 & 5.00& 2.42e-9 & 4.98&  6.87e-10 & 5.00 & 2.42e-9 &4.98\\
\hline
\end{tabular} \label{tab51}
\end{table}

Results listed in Tab \ref{tab51} shows, if the solution is strictly positive, optimal convergence is achieved on sufficient fine grids.

\subsection{a case with discontinuity in Cartesian coordinate system}
The following problem, whose exact solution is always no less than $0$, was used to test the positivity-preserving limiter.
\begin{subequations}
\begin{eqnarray}
&\frac{\partial u}{\partial t}+\frac{\partial u}{\partial x}=0, ~~~~~~x \in [-1,1] \\
& u(x, t=0) = \begin{cases}
\exp\left\{\frac{-\ln 2}{36 \times 0.005^2}(x+0.7)^2\right\},&-0.8 \leq x \leq -0.6,\\
1,&-0.4 \leq x \leq -0.2,\\
1-10|x-0.1|, & 0 \leq x \leq 0.2,\\
\sqrt{1-10^2(x-0.5)^2}, & 0.4 \leq x \leq 0.6,\\
0, & \mbox{otherwise}.
  \end{cases}
\end{eqnarray}
\end{subequations}
with periodic boundary condition.

\begin{figure}[!htp]
\centering
  \subfigure[numerical solutions by WENO5]{
    \label{weno5}
    \includegraphics[width=0.485\textwidth]{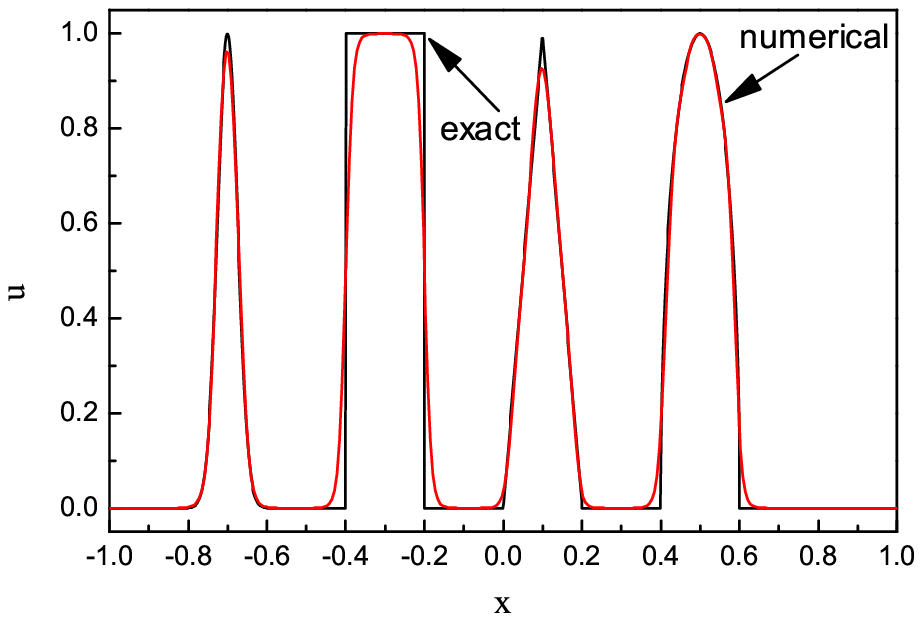}}
  \hfill
  \subfigure[numerical solutions by MUSCL with minmod limiter]{
    \label{weno5re}
    \includegraphics[width=0.485\textwidth]{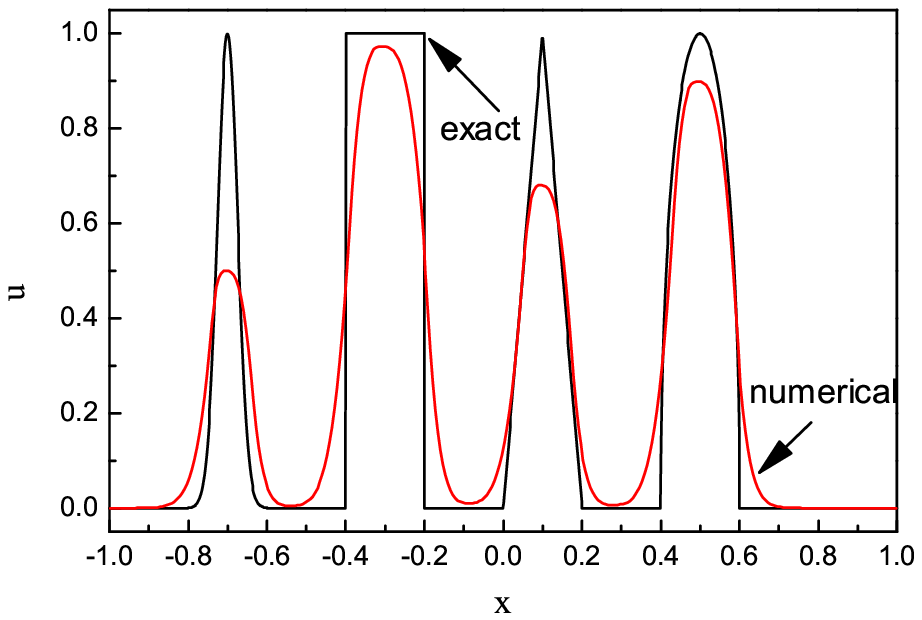}}
  \caption{a comparison of WENO5 with positivity-preserving limiter and MUSCL, computed with 320 points}
\end{figure}

Fig. \ref{weno5} shows the numerical solution of WENO5 with the positivity-preserving limiter. As a comparison, we present the result of MUSCL with minmod limiter in Fig. \ref{weno5re}. Though MUSCL preserves the positivity (c.f. Tab \ref{case2}), however, it is more diffusive, for which we choose a high order WENO scheme as the basis to build our scheme for streamer simulations.

Tab \ref{case2} shows that without the positivity-preserving limiter, the numerical solution by WENO5 violates the positivity. We also remark that, for this example, when the limiter turns on, one would better write the limiter in an equivalent form, $\widehat{h}_{i+\frac{1}{2}}^- = (1- \frac{\overline h_{i} - h_{i+\frac{1}{2}}^-}{\overline h_{i}-q_{\min}}) \overline h_{i}$, to reduce the round-off error.

\begin{table}[!h]
\caption{the minimum of the point values computed with WENO5 and MUSCL, $\triangle t=(\triangle x)^{\frac{5}{3}}$}
\centering
\begin{tabular}{c c c c c}
\hline \hline
\multirow{2}{*}{$\frac{2}{\triangle x}$} & \multirow{2}{*} {exact} &
\multicolumn{2}{c}{5th order WENO} & \multirow{2}{*}{MUSCL} \\
\cline{3-4}
& &\mbox{without limiter}  & \mbox{with limiter}\\
\hline
${80}$ &  0 & -1.94e-2 & 1.94e-07 & 3.15e-3\\
${160}$ & 0 & -7.21e-4 & 3.44e-34 & 1.86e-5\\
${320}$ & 0 &-2.41e-6 & 4.40e-40 & 1.37e-9\\
${640}$ & 0 &-3.86e-7 & 3.78e-80 &3.77e-17\\
\hline
\end{tabular} \label{case2}
\end{table}

\section{Results}
\subsection{In non-attaching gas} 
\begin{figure}[!hbp]
\centering
  \includegraphics[width=.5 \textwidth]{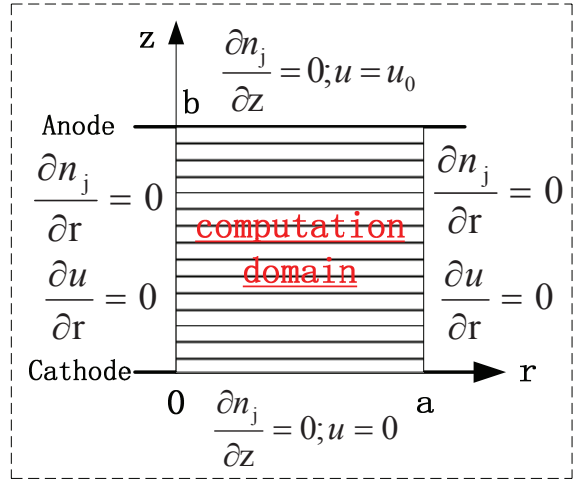}\\
  \caption{the configuration of the discharge simulation}\label{config}
\end{figure}

We first test a double headed streamer discharge simulation in Nitrogen. The configuration is shown in Fig \ref{config}, $U_0=52$ kV, $P= 760$ Torr, $a=b=1.0$ cm and all other coefficients can be found in \cite{dhali1987}. For clarity, we omit the negative ions for non-attaching gases. The initial condition is $n_e=n_p = 10^{14} \exp\left\{-(\frac{r}{0.021})^2-(\frac{z-0.5}{0.027})^2\right\}+ 10^8$ cm$^{-3}$. The time step used is $10^{-13}$ s.

\subsubsection{Comparisons of results by Method I and Method II}
Under same simulation configuration, results computed by the Method I (Strang splitting with exactly solved reaction) or the Method II should agree. In our comparison, the relative difference between A and B is always defined as $\frac{|A-B|}{|B|}$ provided $B\neq 0$.

\begin{figure}[!hbp]
\centering
  \subfigure[electric field along z-axis by Method I]{
    \label{ef-str2}
    \includegraphics[width=0.485\textwidth]{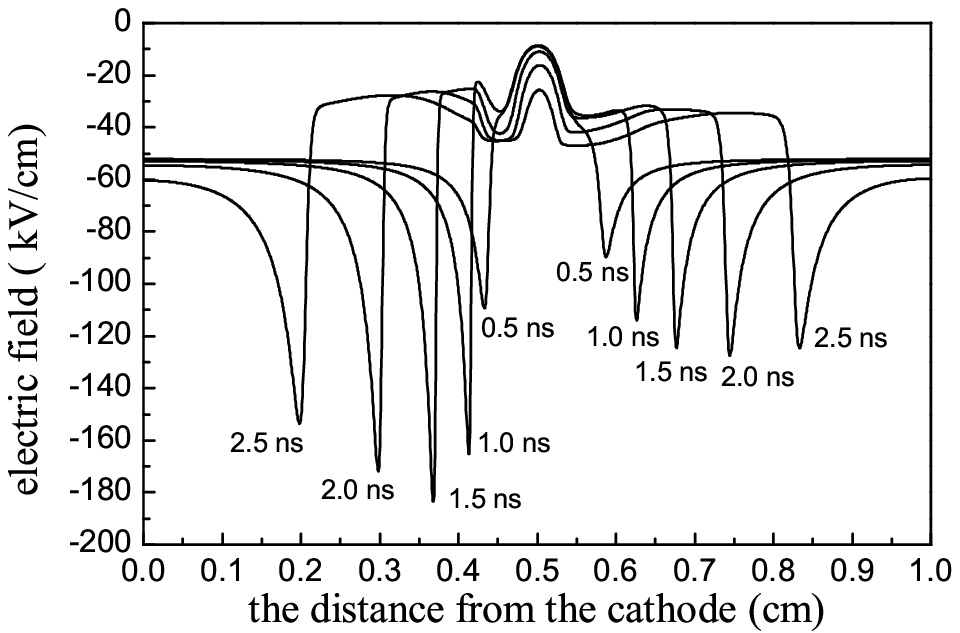}}
  \hfill
  \subfigure[relative difference of electric field along z-axis between Method I and II]{
    \label{re-str-rk}
    \includegraphics[width=0.485\textwidth]{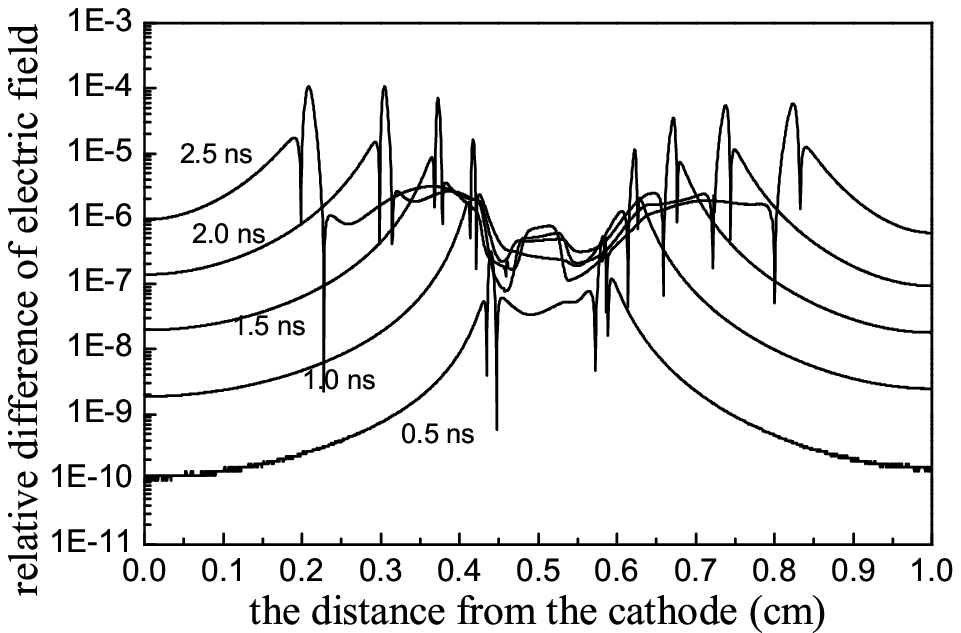}}
  \caption{calculated electric field along z-axis  by the Method I and Method II}
\end{figure}

Fig \ref{ef-str2} shows the electric field along the z-axis obtained by the Method I at different times(1 ns $=10^{-9}$ s) and Fig. \ref{re-str-rk} shows the results obtained by the two methods agree with each other. The electric field is largely enhanced and move towards the opposite electrodes.

\begin{figure}[!h]
\centering
  \subfigure[by Method I at $t=2$ ns]{
    \label{str-t2-dis}
    \includegraphics[width=0.485\textwidth]{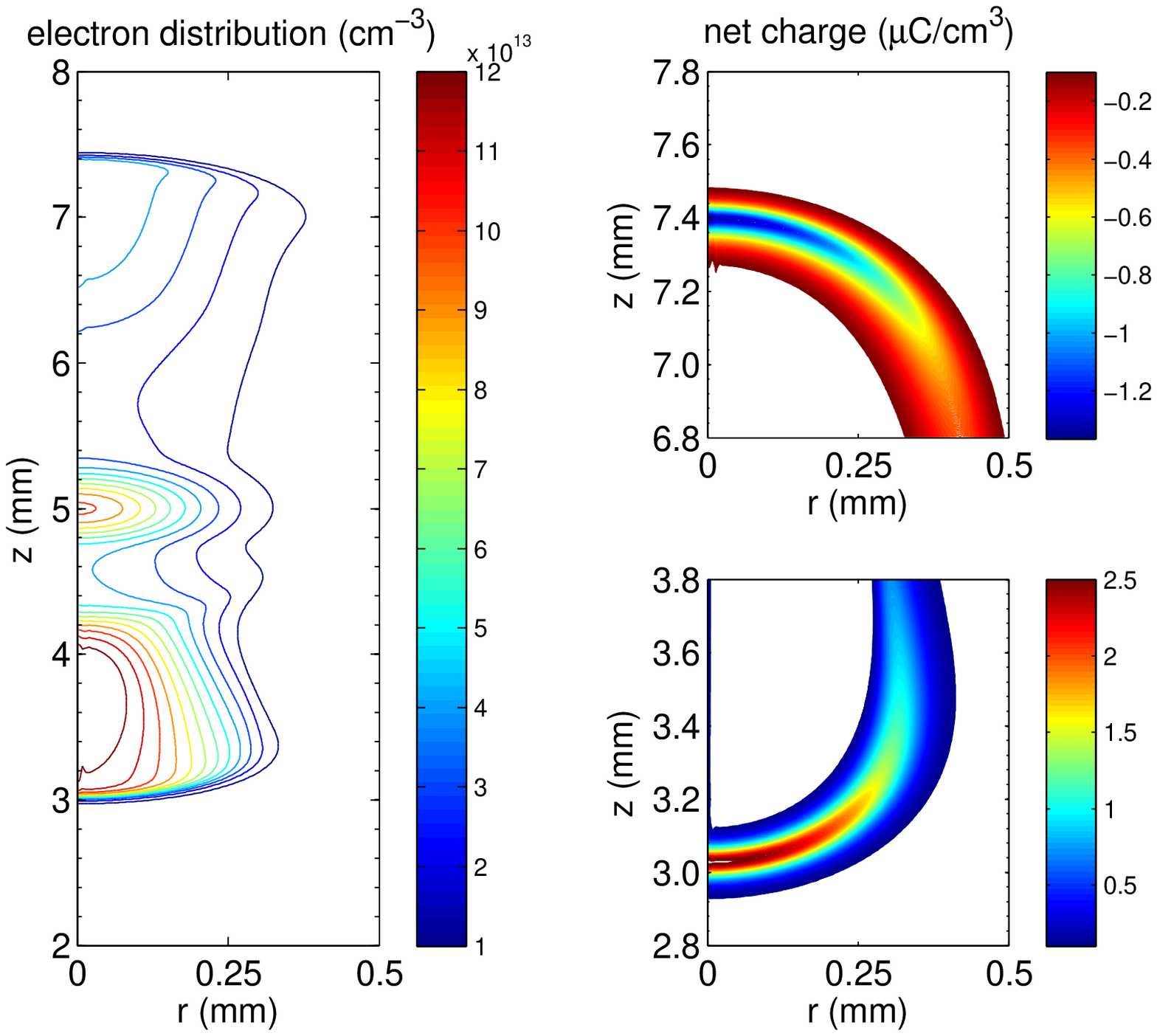}}
  \hfill
  \subfigure[by Method II at $t=2$ ns]{
    \label{rk-t2-dis}
    \includegraphics[width=0.485\textwidth]{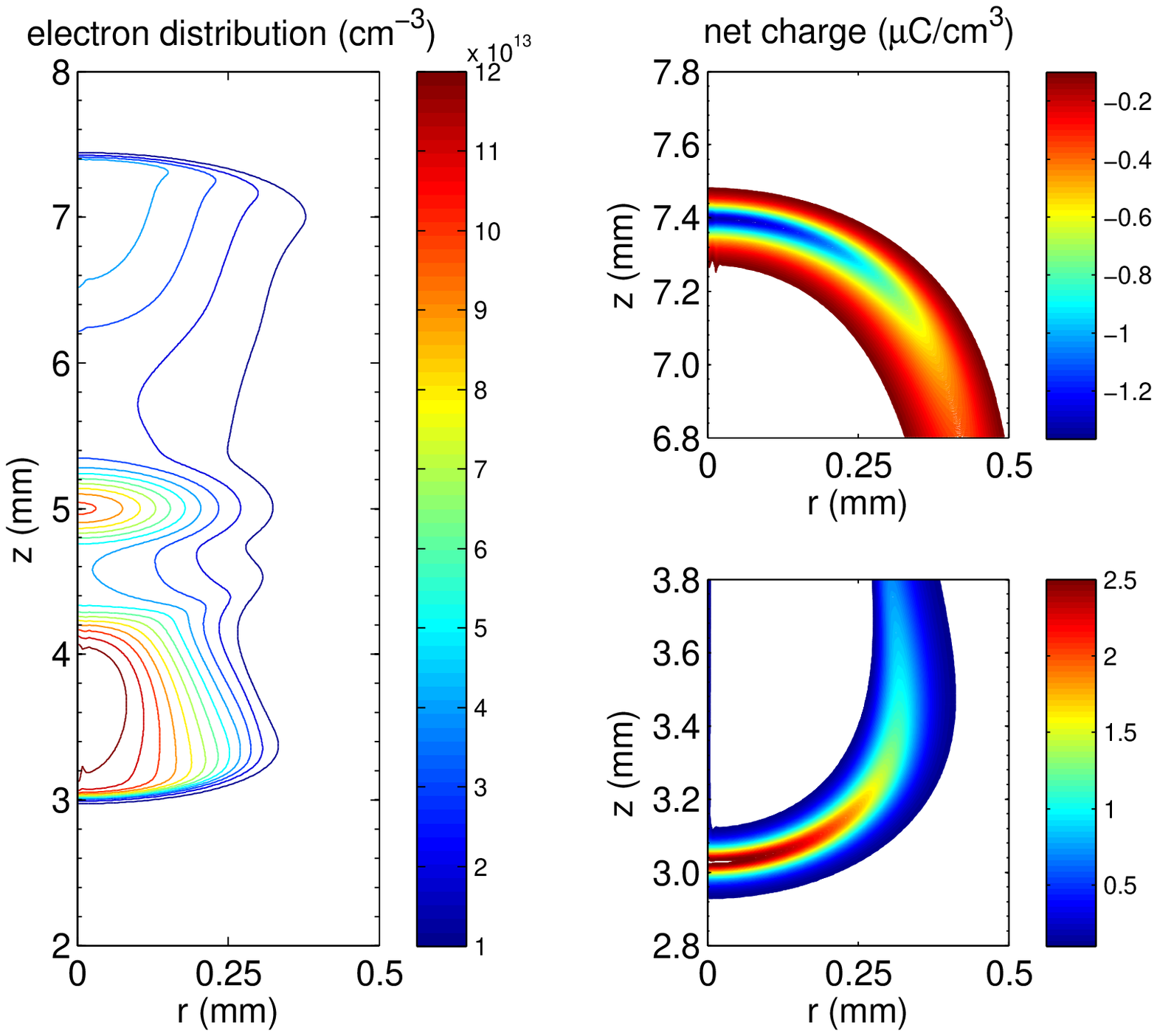}}
     \subfigure[by Method I at $t=2.5$ ns]{
    \label{str-t25-dis}
    \includegraphics[width=0.485\textwidth]{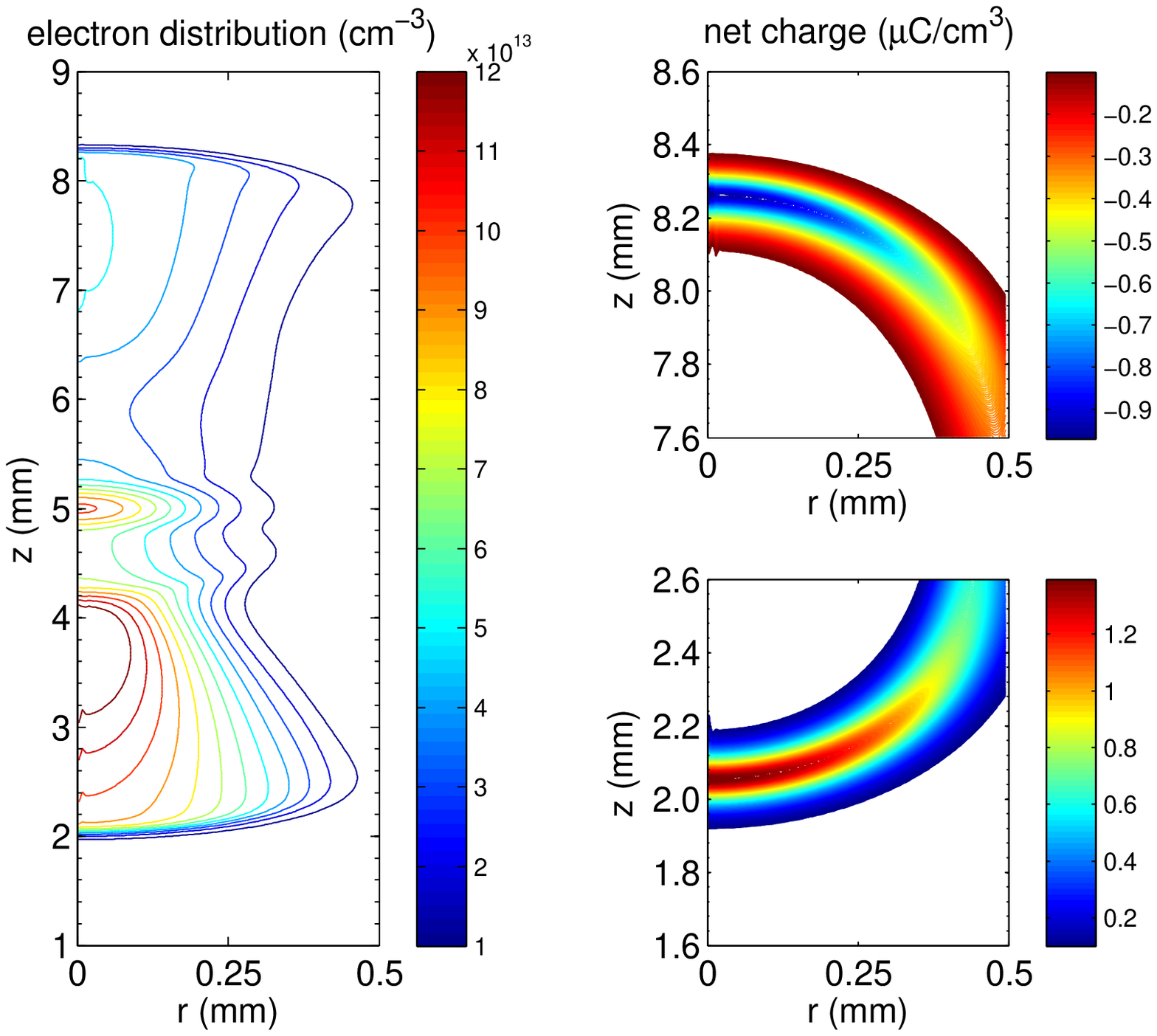}}
  \hfill
  \subfigure[by Method II at $t=2.5$ ns]{
    \label{rk-t25-dis}
    \includegraphics[width=0.485\textwidth]{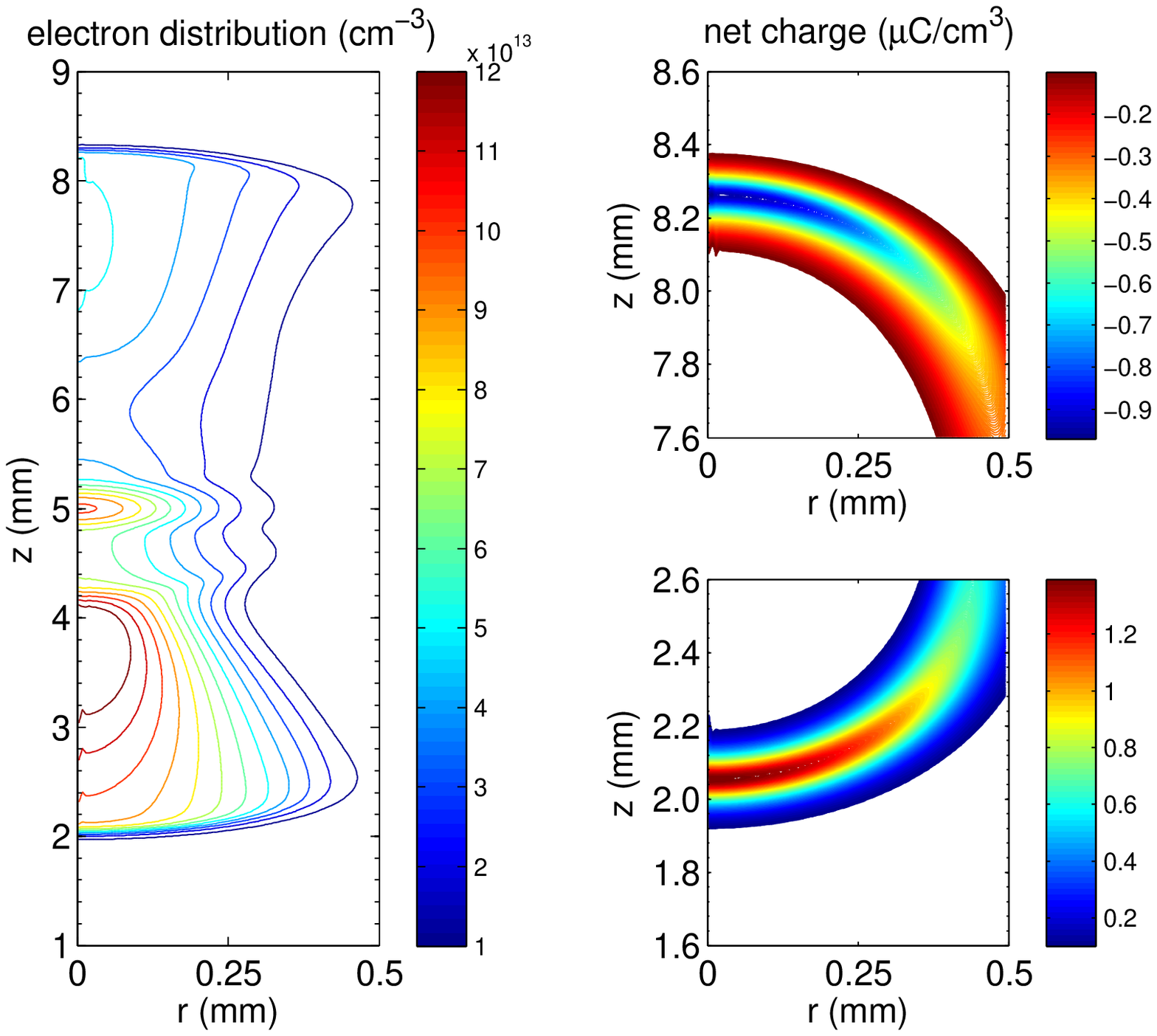}}
  \caption{Computed charge densities by the Method I and Method II at different times}\label{nonattachingdis}
\end{figure}

Fig \ref{nonattachingdis} shows the electron and net charge distributions at different times by the two methods.

\subsubsection{Influence of heavy ions' movements}
\begin{figure}[!htp]
\centering
  \includegraphics[width=0.5\textwidth]{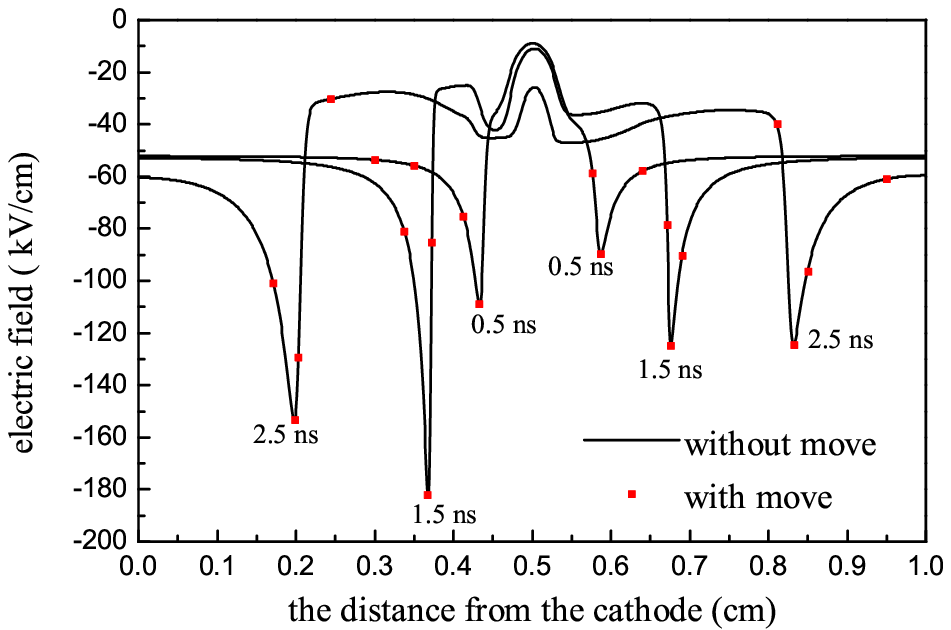}\\
  \caption{the electric field distribution of streamer neglecting the ions' movement}\label{negionsmoveef}
\end{figure}

\begin{figure}[!htp]
\centering
 \subfigure[omitting ions' movement]{
    \label{omit-ion-t225-dis}
    \includegraphics[width=0.485\textwidth]{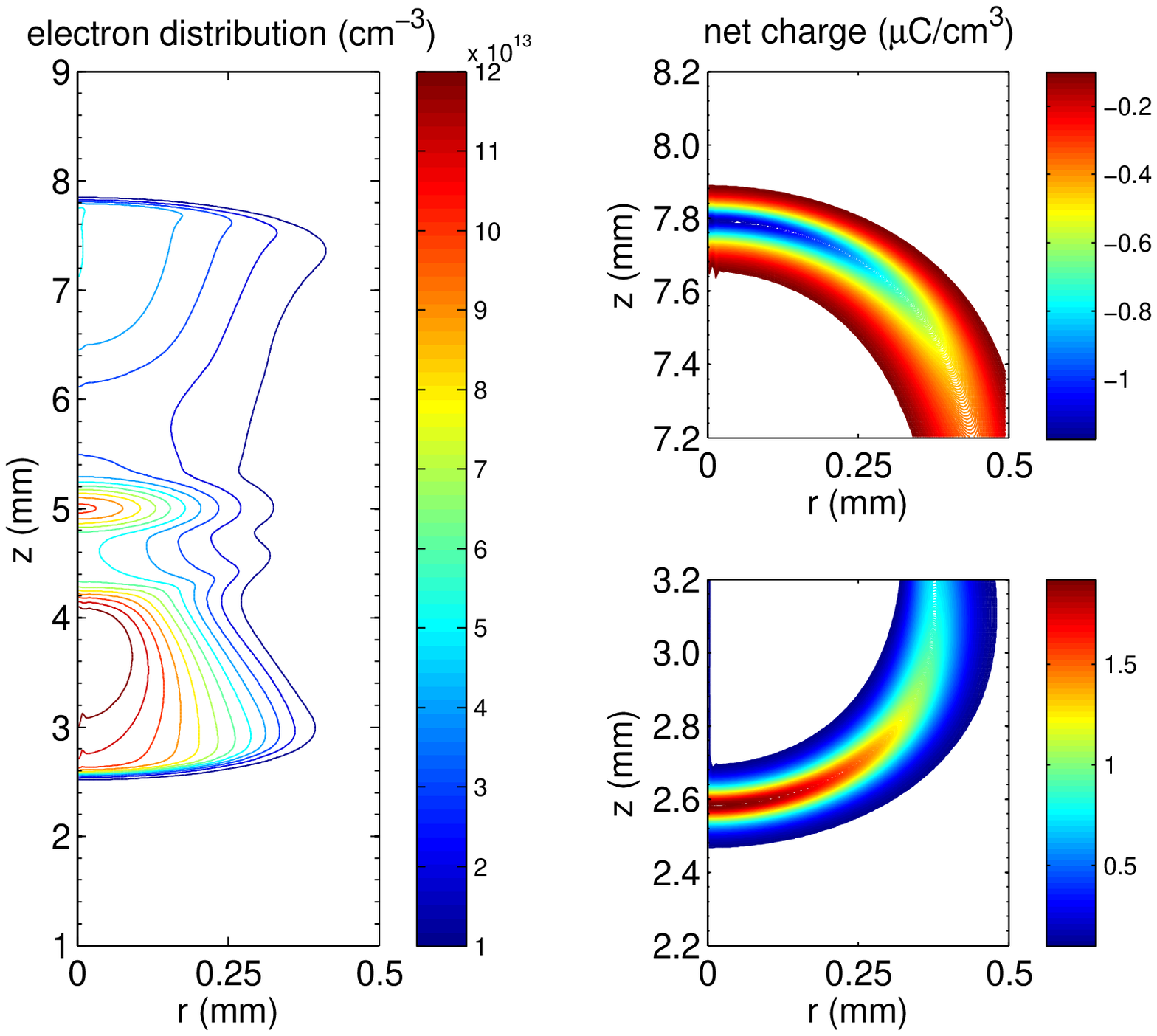}}
     \hfill
     \subfigure[considering ions' movement]{
    \label{whole-t225-dis}
    \includegraphics[width=0.485\textwidth]{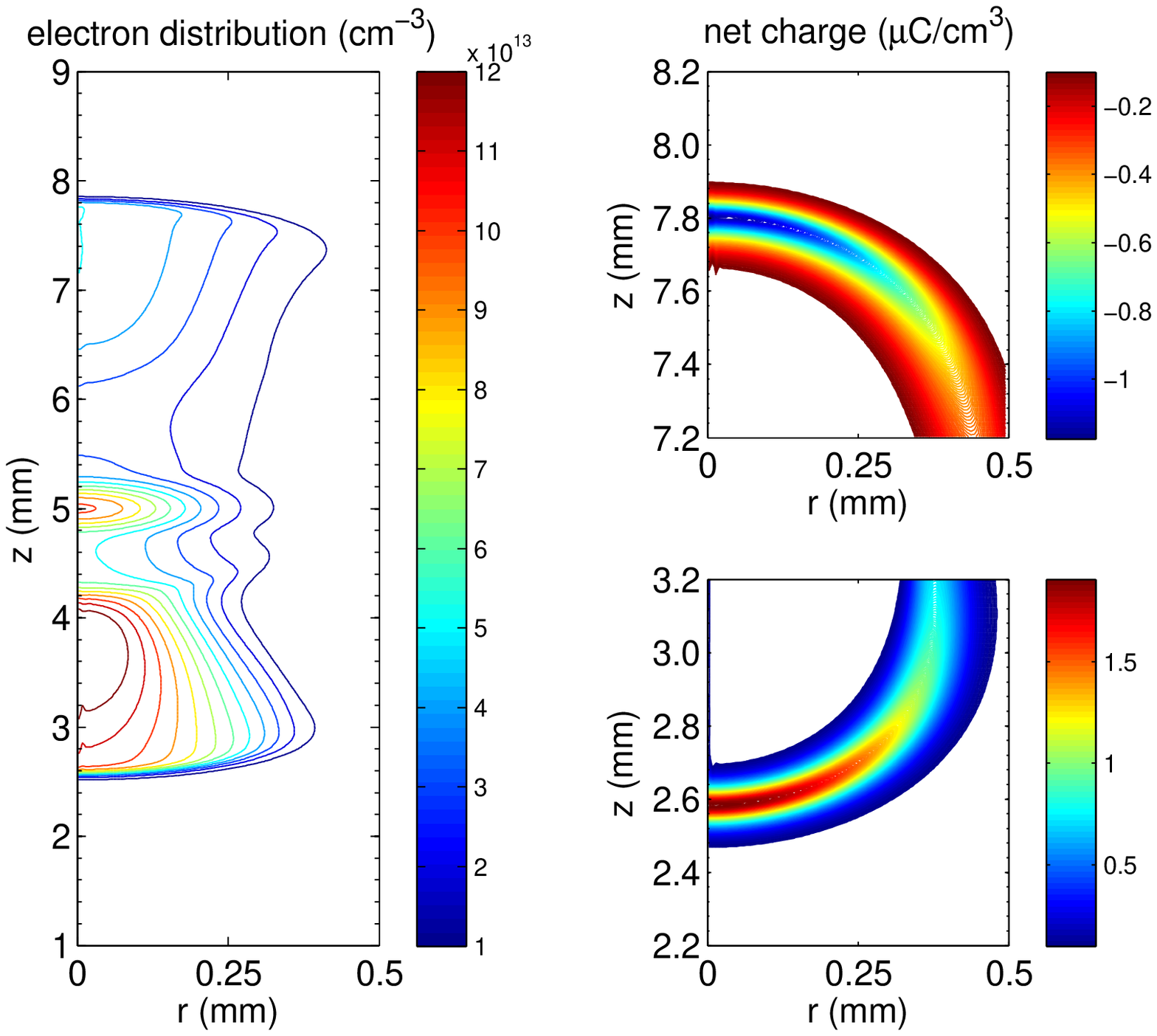}}
    \caption{the charge distribution of streamer neglecting the ions' movement at $t=2.25$ ns}\label{negionsmovecharge}
\end{figure}

The heavy ions drift much more slowly than electrons. If the electrons move 10 cm, the ions move about 1 mm. Compared with the rapid streamer propagations, ions remain almost static. Fig. \ref{negionsmoveef} and Fig. \ref{negionsmovecharge} show, omitting the heavy ions' movements, the electric field along z-axis and the charge distribution almost remain unchanged.

We conclude that the movements of the heavy ions have little effect on the streamer propagation, and can be omitted, which makes Eq (1.2) and Eq (1.3) reduce to ODEs and largely simplifies the model.

\subsection{In attaching gas} We give some simulation results of streamer discharges in SF$_6$. The configuration is same as Fig. \ref{config}.
We choose $N=2.446\times 10^{25}$ m$^{-3}$, $a=b=0.5$ cm, $U=50$ kV and $\triangle t=10^{-13}$ s. The movements of the heavy ions are neglected due to their little influence.

\subsubsection{Comparisons between results with and without the limiter}
First we compare the results obtained with and without the limiter. The initial condition is $n_e=n_p = 10^{14} \exp\left\{-(\frac{r}{0.021})^2-(\frac{z}{0.027})^2\right\}+ 10^4$ cm$^{-3}$ and $n_n=0$. On this occasion, the numerical solution of the electron density will keep positive even without the limiter due to strong background photo-ionizations. However, the positivity-preserving limiter did turn on if the above sufficient conditions were enforced, and cost a little more CPU time than the case without the limiter.

\begin{figure}[!h]
\centering
  \subfigure[densities computed without the limiter.]{
    \label{off-t1-dis}
    \includegraphics[width=0.485\textwidth]{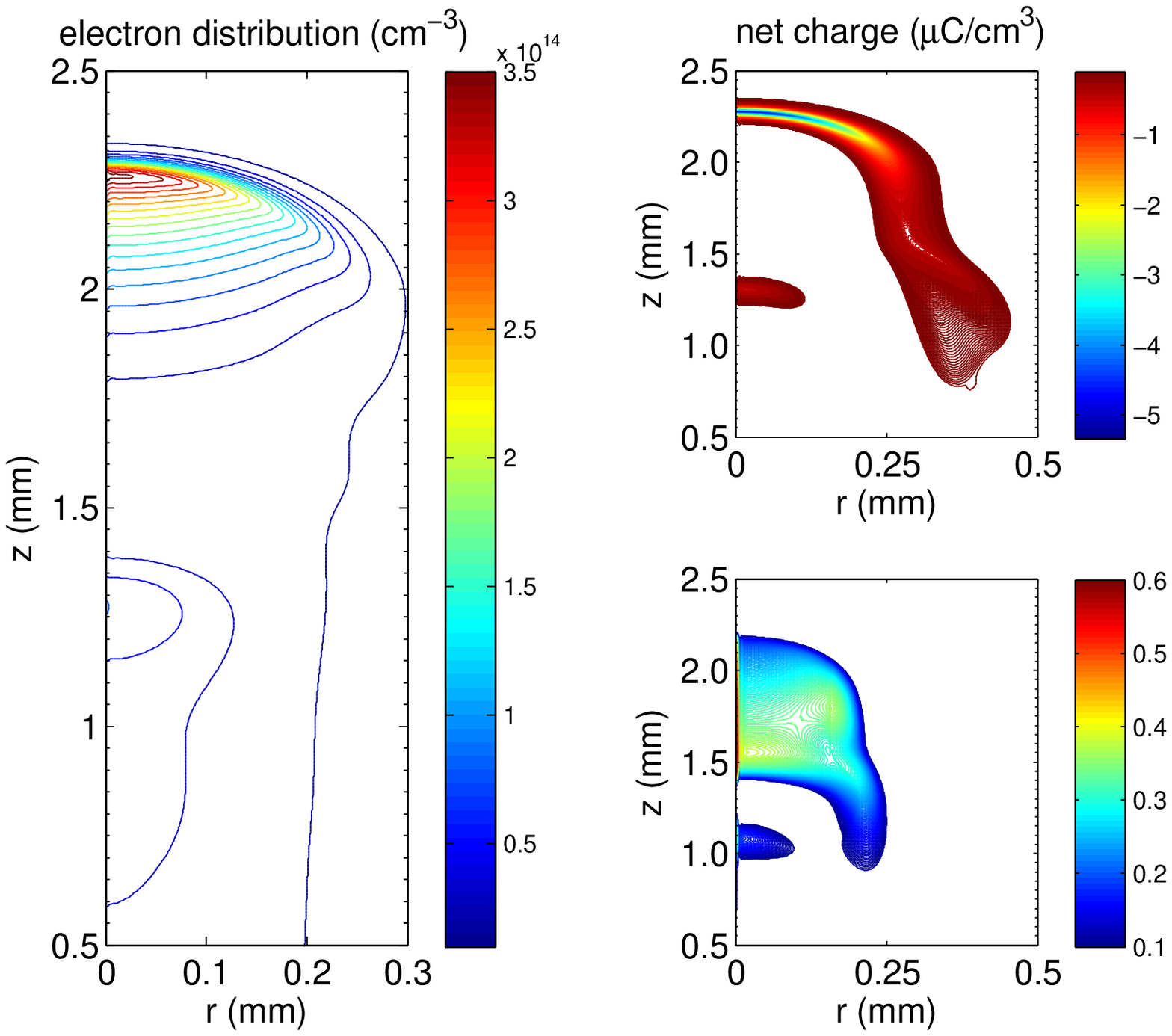}}
  \hfill
  \subfigure[densities computed with the limiter.]{
    \label{on-t1-dis}
    \includegraphics[width=0.485\textwidth]{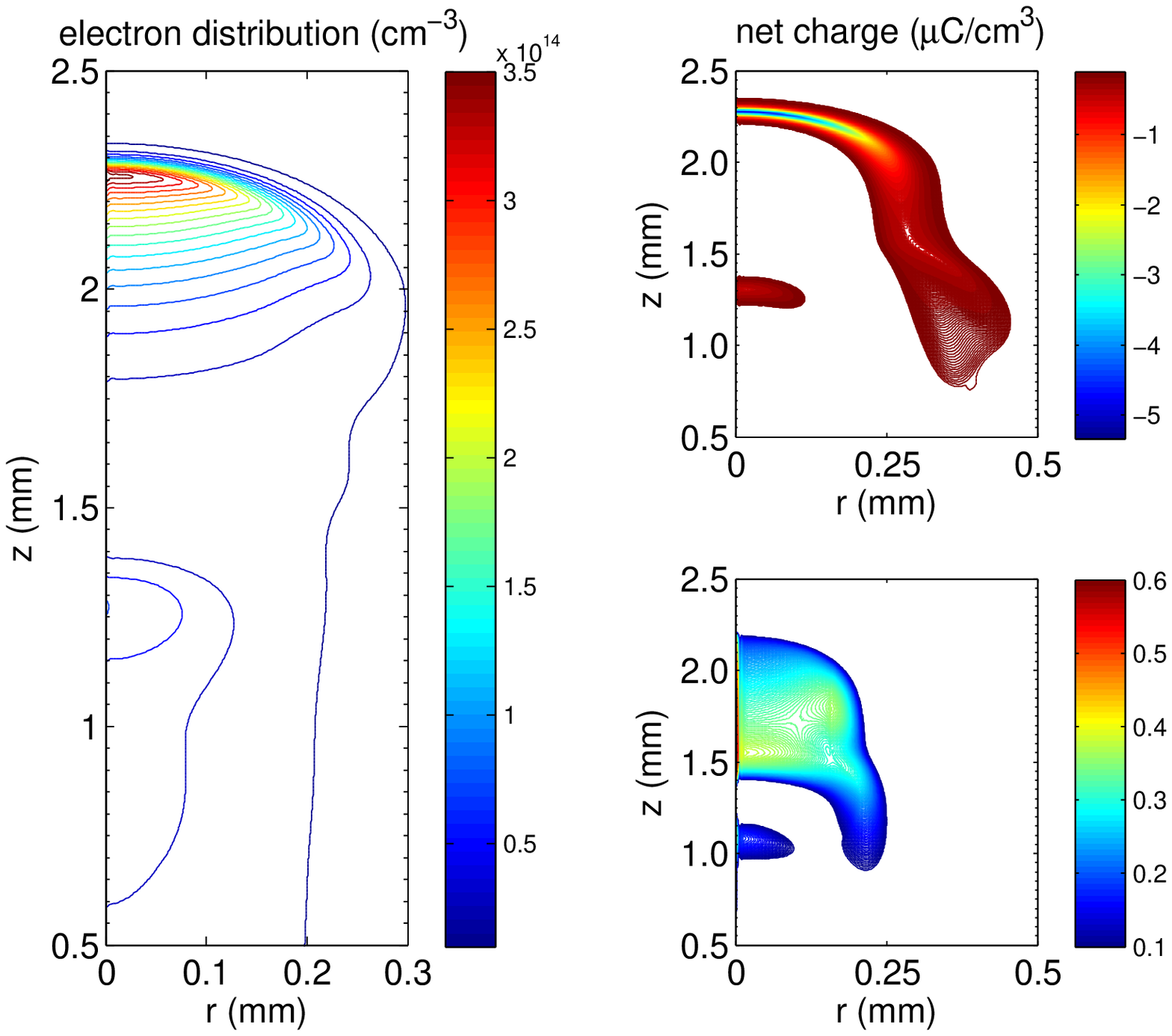}}
  \caption{the particle density distributions obtained with and without the limiter at $t=1$ ns}
  \label{on-off}
\end{figure}

Fig. \ref{on-off} shows the net charge density distributions in the space at $t=1$ ns. The obtained net particle densities with and without the limiter agrees with other and the relative difference is small, e.g., with the limiter, the obtained maximal positive and negative net charge densities are 0.69190 $\mu$C/cm$^{-3}$ and 5.3390 $\mu$C/cm$^{-3}$, respectively; while without the limiter, these values are 0.69206 $\mu$C/cm$^{-3}$ and 5.3390 $\mu$C/cm$^{-3}$, respectively. From Fig. \ref{on-off}, the charge distributions for attaching gas is more complex than those of non-attaching gases. There are both negative net charge area and positive net charge area in the streamer channel: the outer is mainly negative net charge area and  inner is mainly positive net charge area. In addition, the maximal of the negative net charge density is about 10 times larger than that of positive net charge density.

\begin{figure}[!h]
\centering
  \subfigure[charged particle densities along the z-axis, computed without the limiter.]{
    \label{densitycomp1}
    \includegraphics[width=0.485\textwidth]{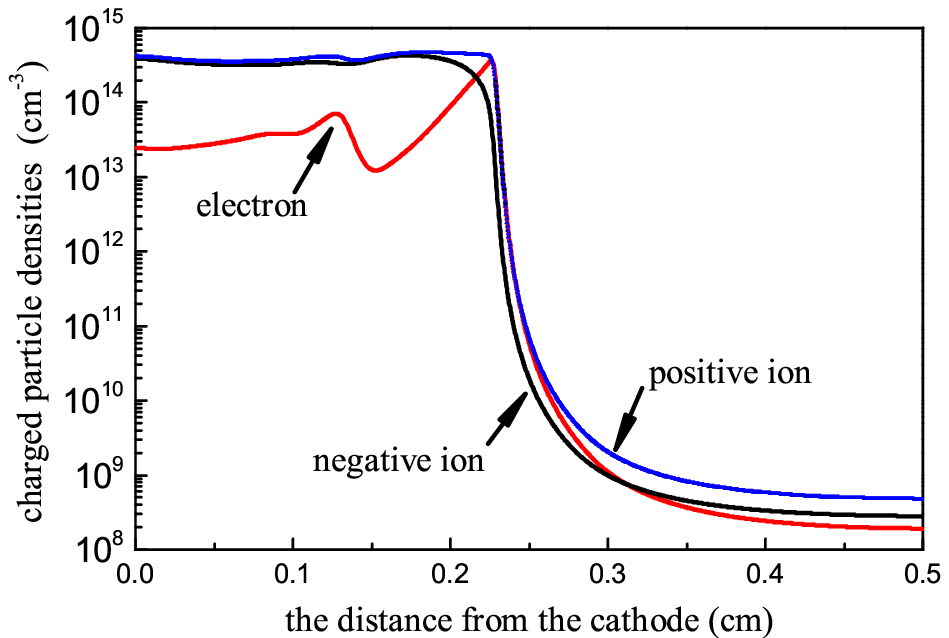}}
  \hfill
  \subfigure[the relative difference of the charged particle densities computed with and without the limiter.]{
    \label{densitycomp2}
    \includegraphics[width=0.485\textwidth]{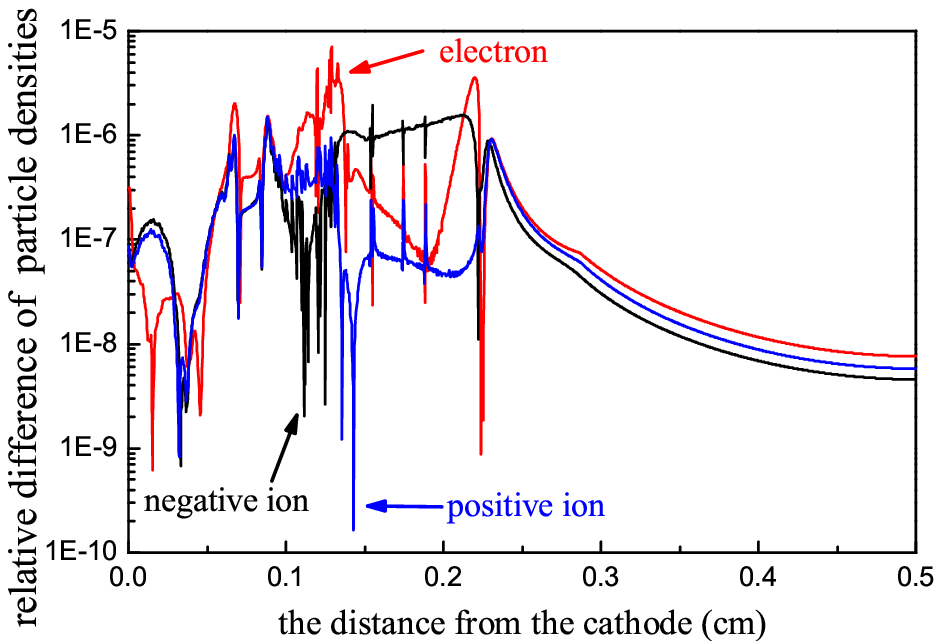}}
  \caption{the particle densities along the z-axis obtained with and without the limiter at $t=1$ ns}
  \label{fig:dencomp}
\end{figure}

Specifically, Fig. \ref{fig:dencomp} shows the space distributions of charged particle densities along the z-axis at $t=1$ ns. The obtained particle densities with and without the limiter agrees with each other. From Fig. \ref{densitycomp1}, due to the attachment, the electron density in the
body of the streamer is reduced by approximately one to two orders of magnitude. However, the ion densities were much larger, approximately one order of magnitude, than those of non-attaching gases, due to the stronger collision ionizations and attachments caused by the higher BDEF.

\begin{figure}[!h]
\centering
  \subfigure[a comparison of electric field along the z-axis for cases with and without the limiter. black solid: with the limiter; red cross: without the limiter.]{
    \label{compare1}
    \includegraphics[width=0.485\textwidth]{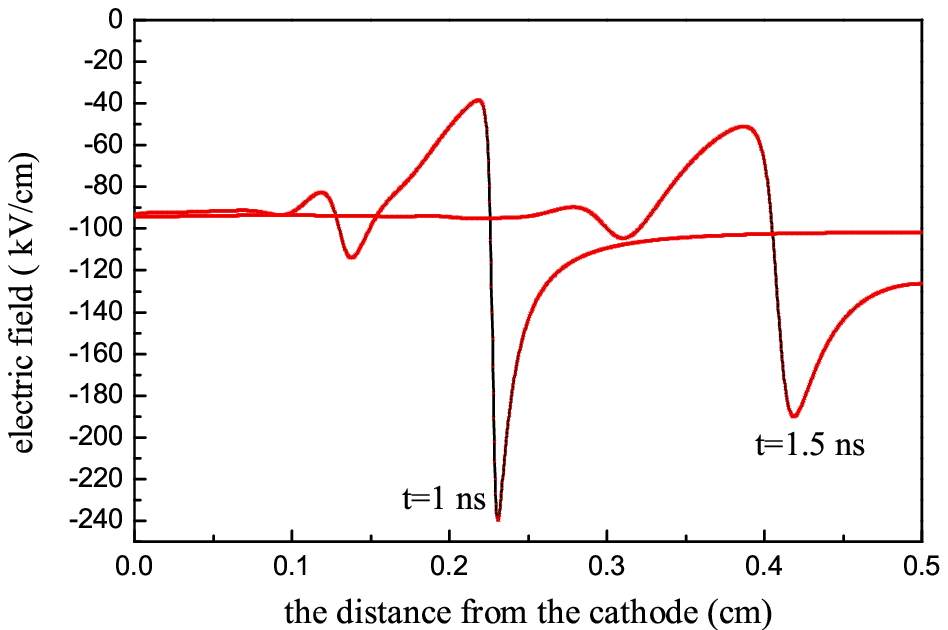}}
  \hfill
  \subfigure[the relative difference of the electric field strength along the z-axis for cases with and without the limiter. black line: $t=1$ ns; red line: $t=1.5$ ns]{
    \label{compare2}
    \includegraphics[width=0.485\textwidth]{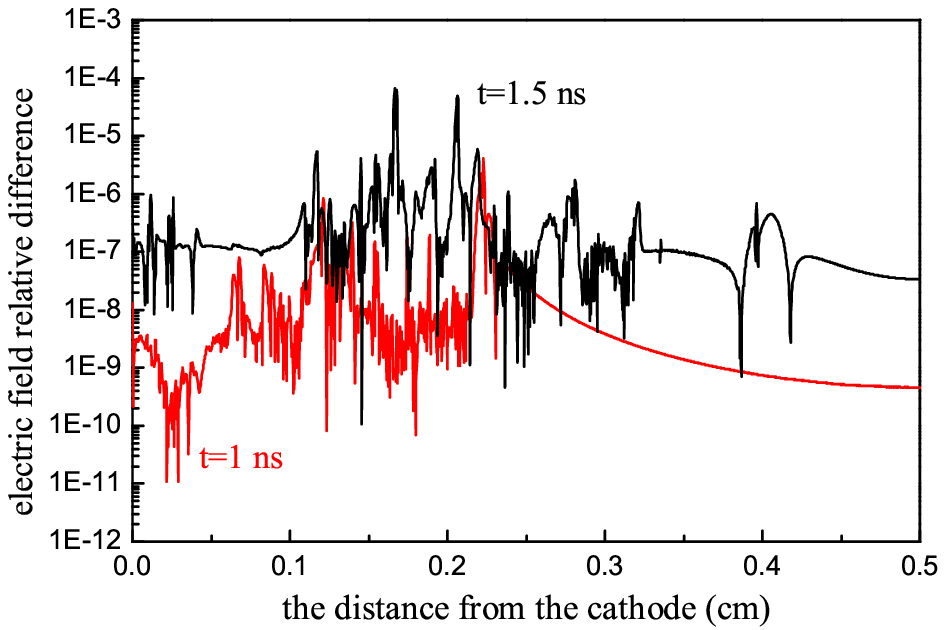}}
  \caption{comparison of electric field along the z-axis with and without the limiter}
  \label{fig:pplcomp}
\end{figure}

Fig. \ref{fig:pplcomp} shows the electric field along the z-axis. The results computed with or without the limiter agrees with each other and the relative difference is small. From Fig. \ref{compare1}, the electric field behind the streamer front was close to the value which makes $\alpha=\eta$ (below we name it as balance electric field(BEF)). For non-attaching gas, e.g., N$_2$, BEF is much smaller than the breakdown electric field (BDEF) and the electric field behind the streamer front may be much above it, which is different from attaching gases.

From the comparisons, we conclude that the positivity-preserving limiter does not change the exact results.

\subsubsection{Results for cases without background photo-ionization}
Secondly, we simulate a negative streamer that develops without any background photo-ionization. The initial condition is $n_e=n_p = 10^{14} \exp\left\{-(\frac{r}{0.021})^2-(\frac{z}{0.027})^2\right\}$ cm$^{-3}$ and $n_n=0$. On this occasion, without positivity-preserving limiter, the WENO finite difference scheme failed to give non-negative charged particle densities. In our simulation, the positivity-preserving limiter turned on at each time step.

\begin{figure}[!h]
\centering
\includegraphics[width=0.485\textwidth]{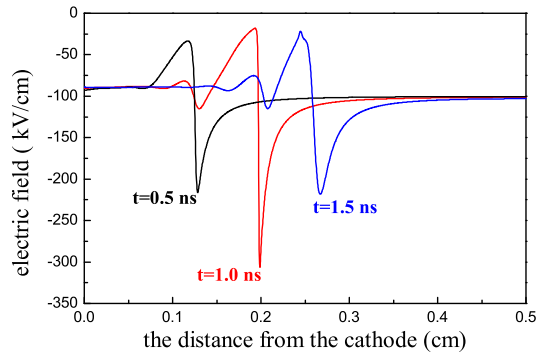} \\
  \caption{the particle densities along the z-axis computed with the limiter at different times (without photo-ionization)}
  \label{bg0ef}
\end{figure}

 Fig. \ref{bg0ef} shows the electric field along the z-axis at different times. Compared with Fig. {\ref{compare1}}, without the photo-ionization, the anode-directed streamer develops more slowly. However, the maximal electric field is nearly 30 percent larger at $t=1$ ns.

\begin{figure}[!h]
\centering
  \subfigure[$t=0.5$ ns.]{
    \label{bg0-t05}
    \includegraphics[width=0.485\textwidth]{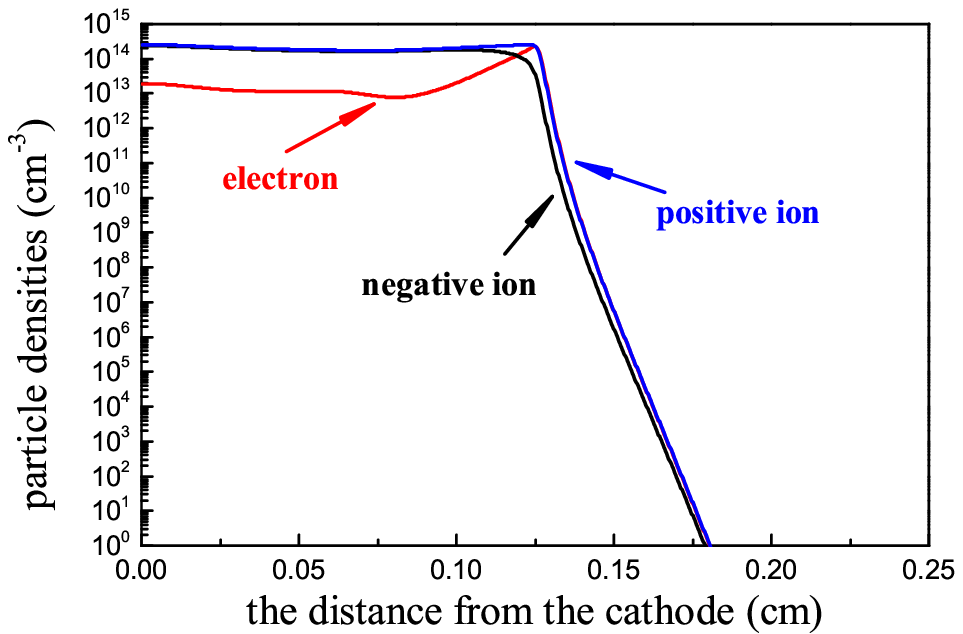}}
  \hfill
  \subfigure[$t=1$ ns]{
    \label{bg0-t10}
    \includegraphics[width=0.485\textwidth]{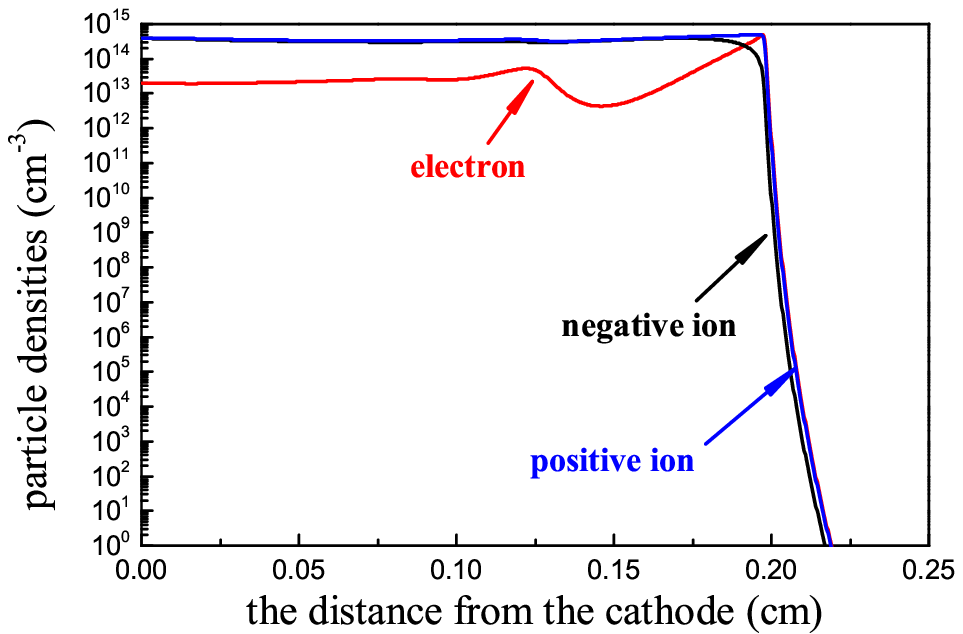}}
  \caption{the particle densities along the z-axis computed with the limiter at different times (without photo-ionization)}
  \label{bg0}
\end{figure}

Fig. \ref{bg0} shows the particle densities along the z-axis at different times.
Similarly to the results shown in the last section, the electron density in the streamer channel is also reduced by approximately one to two orders of magnitude due to the attachment and the negative ion density is of the same order of positive ion density.

\begin{figure}[!h]
\centering
  \subfigure[$t=0.5$ ns.]{
    \label{bg0-t05-dis}
    \includegraphics[width=0.485\textwidth]{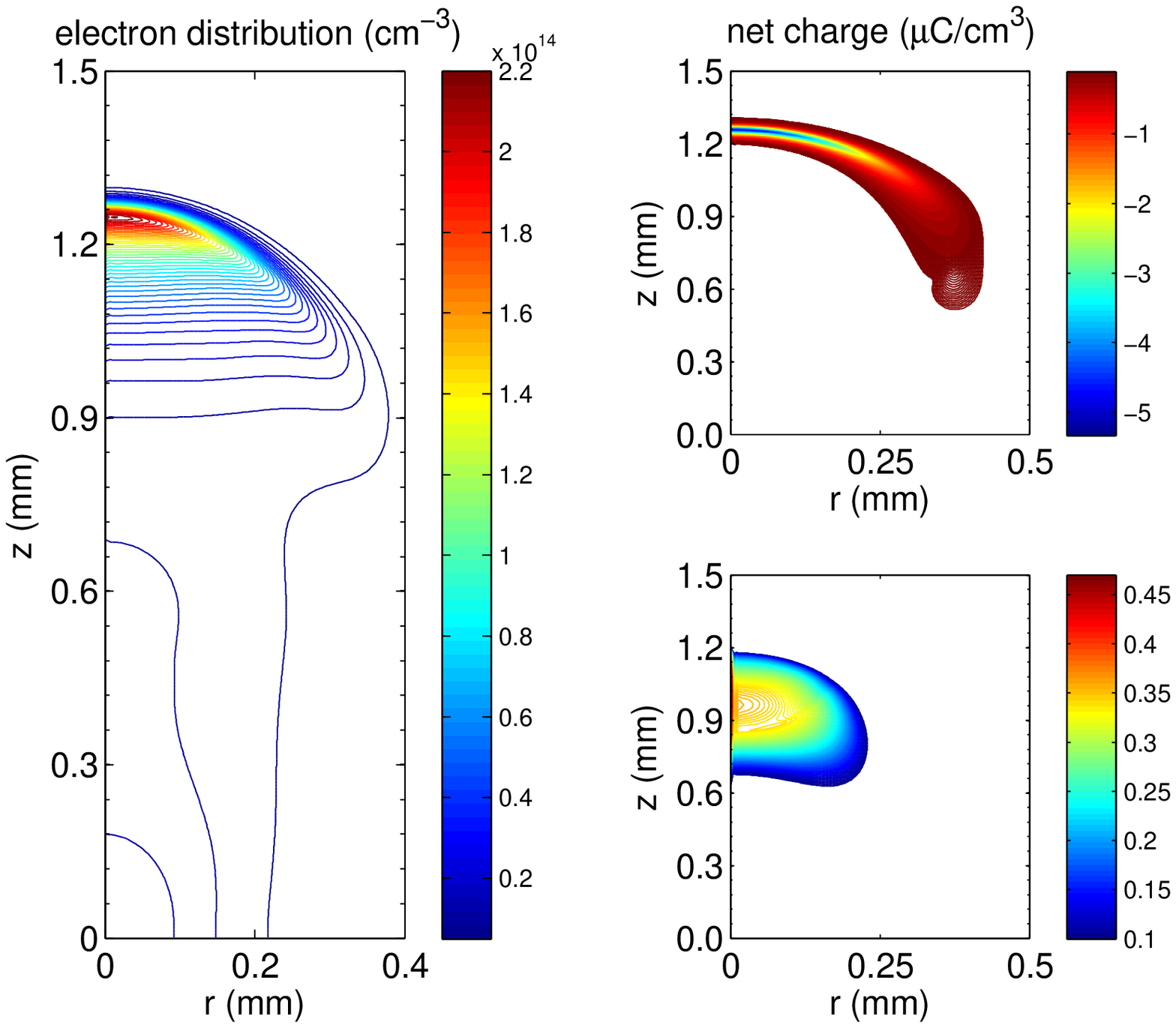}}
  \hfill
  \subfigure[$t=1$ ns]{
    \label{bg1-t1-dis}
    \includegraphics[width=0.485\textwidth]{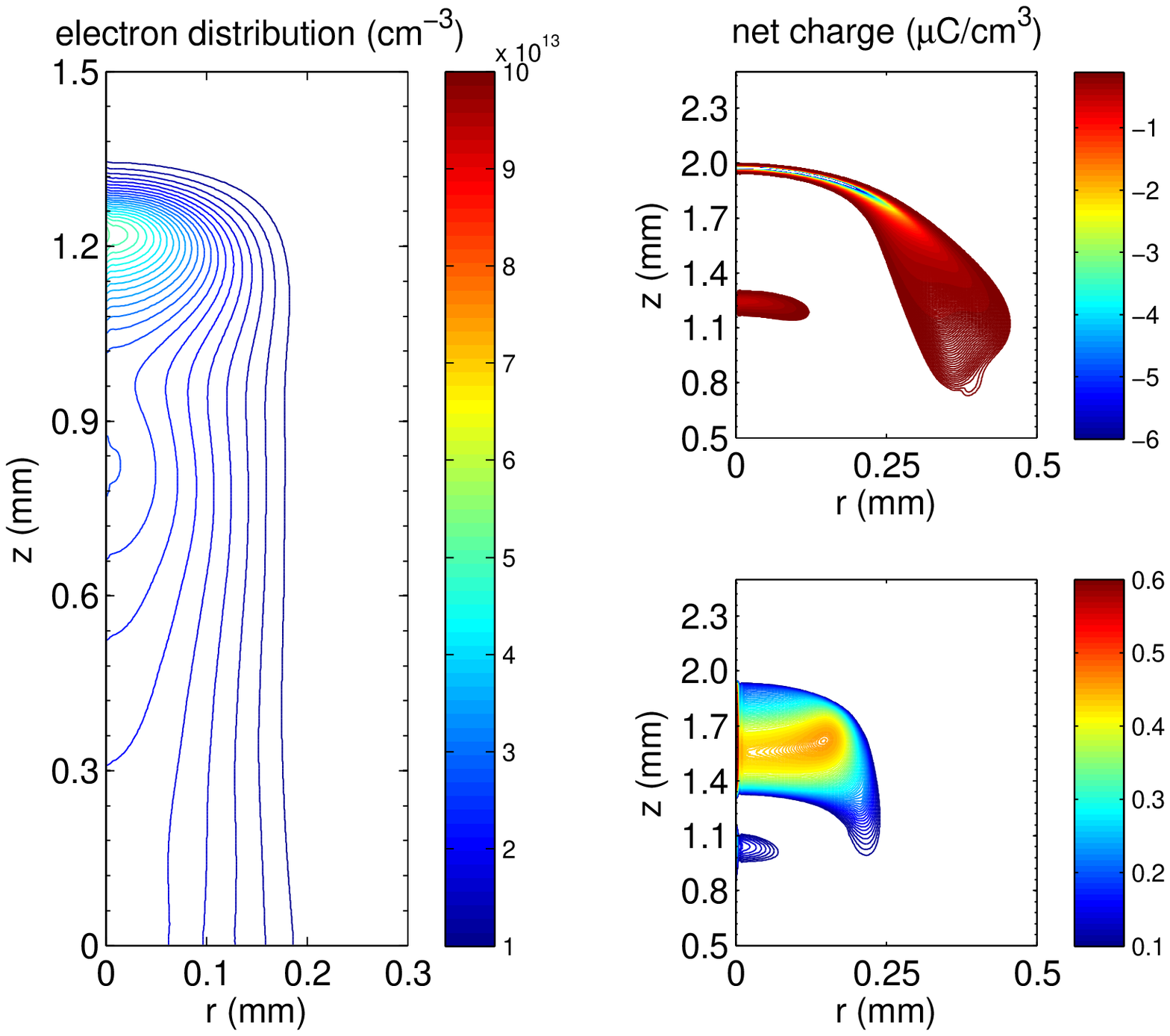}}
  \caption{the particle density distribution computed with the limiter at different times (without photo-ionization)}
  \label{bg0den}
\end{figure}

Fig. \ref{bg0den} show the particle densities at different times. Besides that the streamer develops much more slowly, the shape of net charge densities distribution profile are similar to the case of $10^4$ background photo-ionizations. Both negative net charge area and positive net charge area are in the streamer channel, where the positive net charge mainly concentrates in the middle area of the channel and the negative net charge mainly surrounds the positive net charge area.

In the view of positivity-preserving property, the cases without photo-ionizations in attaching gases is the worst situations for a simulation algorithm.
The proposed scheme does work and preserve the positivity of the densities.

\section{Conclusion}
This paper proposed a finite difference scheme for the numerical simulation of streamer discharges in non-attaching and attaching gases, which guarantees the positivity of particle densities. It uses the WENO finite difference scheme together with the positivity-preserving limiter proposed by Zhang and Shu, and Strang splitting as well. The positivity-preserving property is provable under a stricter CFL restriction. Different from the slope limiters like $minmod$, $Superbee$, the positivity-preserving limiter can be turned off when the positivity is not violated, hence it would not kill the accuracy at regions with positive values. Numerical simulations of streamer discharges in a non-attaching gas (N$_2$) and attaching gas (SF$_6$) are given to illustrate the effectiveness of the scheme.

Positivity-preserving streamer discharge simulation schemes on unstructured grids and simulations with more accurate photo-ionization models are under working.

\section*{Acknowledgement}
Dr. Zhang, once with Brown University, now with MIT, is greatly appreciated for the helpful discussion on the positivity-preserving limiter. The first author appreciates the mathematicians home and abroad who helped him greatly when he was a PhD student
, especially Prof. Weizhu Bao at National University of Singapore, Prof. Tiegang Liu at Beihang University, Prof. Yingjie Liu at Georgia Institute of Technology, Prof. Chi-Wang Shu at Brown University, and Prof. Huazhong Tang at Peking University (The names are listed in alphabetical order). In addition, there would be a long name list if fully enumerated. As the Chinese saying goes, D${\grave a}$ ${\bar E}$n B${\grave u}$ Y${\acute a}$n Xi${\grave e}$.

The reviewers are greatly appreciated for their careful reading and helpful comments which makes a great improvement of the manuscript.

This work is supported by National Basic Research Program of China (973 program) under grant 2011CB209403 and National Natural Science Foundation of China under grant 51207078.

\section*{Appendix}
The finite volume scheme used in some papers on streamer discharge simulations is given (for clarity, a uniform mesh is used and the diffusion and $\frac{\partial F}{\partial z}$ terms are omitted):
\begin{equation}
  \frac{\mbox{d}u}{\mbox{d}t}=\frac{1}{r_i \triangle r}\left(r_{i-\frac{1}{2}} \widehat F_{i-\frac{1}{2},j}-r_{i+\frac{1}{2}} \widehat F_{i+\frac{1}{2},j}\right) + S_{i,j}, \label{ap}
\end{equation}
where the numerical flux $\widehat F_{i\pm\frac{1}{2},j}$ is constructed using a slope limiter, e.g, the $minmod$ limiter, and $r_i > 0$. Assume $F(u)=vu$, which is the cases in streamer discharge simulations, and $F'(u)=v^+ \ge 0$, then
\begin{equation}
  \widehat F_{i+\frac{1}{2},j}=v_{i+\frac{1}{2},j}^+\left[u_{i,j}+0.5\Phi(\theta_{i,j})(u_{i+1,j}-u_{i,j})\right],
\end{equation}
in which $\theta_{i,j}=\frac{u_{i,j}-u_{i-1,j}}{u_{i+1,j}-u_{i,j}},\mbox{~~}\Phi(\theta)=\max\left(0,\min(1,\theta)\right)$.

Assume $S_{i,j}=0$, $v_{i+\frac{1}{2},j}=1$, $u_{i,j}^0=i+1$, then at $r_i=0.5 \triangle r$, $i=0$,
\begin{equation}
  u_{i,j}^1=u_{i,j}^0+\frac{\triangle t}{\triangle r}\left(\frac{r_{i-\frac{1}{2}}}{r_i} \widehat F_{i-\frac{1}{2},j}-\frac{r_{i+\frac{1}{2}}}{r_i}\widehat F_{i+\frac{1}{2},j}\right)
  =1-3\frac{\triangle t}{\triangle r}.
\end{equation}

In general, MUSCL scheme with $minmod$ limiter is positivity-preserving, under the CFL condition $\frac{\max|v|\triangle t}{\triangle r}\le \frac{2}{3}$ \cite{shma}. However, scheme Eq (\ref{ap}) is not positivity-preserving under this CFL condition near $r=0$, even if the source term is non-negative. Other limiters, e.g., $Koren$, $Superbee$, have similar problems if the reconstruction is based on the physical variable $u$ rather than the conservative variable $ru$.

\end{document}